\documentclass[12pt]{article}
\usepackage{epsfig,float,cite}
\newcommand{\spa}{\phantom{$\Big($}}
\newcommand{\nn}{\nonumber}
\newcommand{\bea}{\begin{eqnarray}}
\newcommand{\eea}{\end{eqnarray}}
\def \branch{{\cal B}}
\def \beq{\begin{equation}}
\def \eeq{\end{equation}}
\def \branch{{\cal B}}

\def \gev{{\hbox{GeV}}}

\def \cl#1{{#1\%\ \mathrm{C.L.}}}

\def \nn{\nonumber}

\def \bit{\begin{itemize}}
\def \eit{\end{itemize}}

\def \D{\Delta}
\def \g{\gamma}
\def \G{\Gamma}
\def \d{\delta}

\def \r{\rho}

\def \S{\Sigma}

\def\plb#1#2#3{    {\it Phys. Lett. }{\bf B #1} (#2) #3}

\def\prd#1#2#3{    {\it Phys. Rev. }{\bf D #1} (#2) #3}

\def \ds{\displaystyle}

\begin{document}
\begin{titlepage}


\begin{flushright}
CERN-TH/2003-308\\
SLAC-PUB-10279\\
ZU-TH 22/03
\end{flushright}

\centerline{\Large\bf  Untagged $\bar B\to X_{s+d}\, \gamma$ CP asymmetry}
\vspace*{0.2cm}
\centerline{\Large\bf  as a probe for new physics}

\vspace*{1cm}
\centerline{{\large\bf  Tobias Hurth,$^{a,b}\,$\footnote{~Heisenberg Fellow.}
 Enrico Lunghi,$^{c}$ and Werner Porod$^{c}$             }}
 
\vspace{0.1cm}

\begin{center}
{\it ${}^a$~Theoretical Physics Division, CERN, CH-1211 Geneva 23, Switzerland}\\

\vspace{0.3cm}

{\it ${}^b$~SLAC, Stanford University, Stanford, CA 94309, USA}\\

\vspace{0.3cm}

{\it ${}^c$~Institute for Theoretical Physics, University of Zurich, CH-8057 
 Zurich, Switzerland}\\
\end{center}

\medskip 

\begin{abstract}
The direct CP asymmetry in the untagged inclusive channel $\bar B
\rightarrow X_{s+d}\, \gamma$ provides a strict test of the standard
model. It has been shown beyond the partonic level that this asymmetry
is negligibly small thanks to U-spin relations and to the unitarity of
the CKM matrix. In the present paper we investigate this relation
beyond the SM; in particular, we analyse to which extent deviations
from this prediction are possible in supersymmetric scenarios. We
analyse the minimal flavour violation scenario, including $\tan
\beta$-enhanced terms and using the complete two-loop renormalization
group running. Our analysis fully takes into account also the EDM
constraints on the supersymmetric phases. We investigate possible
correlations between the tagged and the untagged CP asymmetries and
the indirect sensitivity of the latter to the $\bar B \rightarrow X_d
\gamma$ CP asymmetry.

Furthermore, we derive general model-independent formulae for the
branching ratios and CP asymmetries for the inclusive $\bar B
\rightarrow X_d\, \gamma$ and $\bar B \rightarrow X_s\, \gamma$ modes,
and update the corresponding SM predictions. We obtain:
\bea 
\branch [\bar B \to X_s \gamma]
& = & ( 3.61 \,\, \left. {}^{+0.24}_{-0.40} \right|_{m_c / m_b}
                     \pm 0.02_{\rm CKM} \pm 0.24_{\rm param.} \pm 0.14_{\rm scale} ) \times 10^{-4}   \,, \nn \\
A_{\rm CP} [\bar B \to X_s \gamma]
& = & 
( 0.42 \,\,\left. {}^{+0.08}_{-0.08} \right|_{m_c / m_b}\, 
 \left. \pm 0.03_{\rm CKM} {}^{+0.15}_{-0.08} \right|_{\rm scale} ) \; \% \,.\nn  
\eea
\end{abstract}

\end{titlepage}

\section{Introduction}

The CKM prescription of CP violation with one single phase is very
predictive. It was proposed already in 1973~\cite{CKM72}, before the
experimental confirmation of the existence of the second family.
Before the start of the $B$ factories, the neutral kaon system was the
only environment where CP violation had been observed.  It has been
difficult to decide if the CKM description of the standard model (SM)
really accounted quantitatively for the CP violation observed in the
kaon system, because of the large theoretical uncertainties due to
long-range strong interactions. The rich data sets from the
$B$-factories now allow for an independent and really quantitative
test of the CKM-induced CP violating effects in several independent
channels. Within the golden $B$ mode $B_d \rightarrow J/\psi K_S$ the
CKM prescription of CP violation has already passed its first
precision test; in fact, the measured CP violation is well in
agreement with the CKM prediction~\cite{BelleCP,BabarCP}.

Nevertheless, there is still room for non-standard CP phases. An
additional experimental test of the CKM mechanism is provided by the
mode $B_d \rightarrow \Phi K_S$. This mode is induced at the loop
level only and, therefore, it is much more sensitive to possible
additional sources of CP violation than the tree-level-induced decay
$B_d \rightarrow J/\psi K_S$. However, the poor statistics does not
allow to draw final conclusions yet~\cite{BelleCP2,BabarCP2}. Direct
CP asymmetries in loop-induced $\Delta F = 1$ modes allow for
additional precision tests of the mechanism of CP
violation. Currently, these decays are less probed than $\Delta F =2$
transitions. However, very precise measurements of direct CP
asymmetries in inclusive rare $B$ decays, such as $b \rightarrow s$ or
$b \rightarrow d$ transitions, will be possible in the near future and
they are the focus of the present paper.

Within the SM the direct CP asymmetry in the inclusive decay $\bar B
\rightarrow X_s \gamma$ is expected to be below $1 \%$. This is a
consequence of three suppression factors: 
({\it i}) Direct CP violation requires at least two interfering
contributions to the decay rate with different strong and weak phases;
thus, an $\alpha_s$ factor is needed in order to generate a strong
phase.  
({\it ii}) This interference receives a CKM suppression of order
$\lambda^2$. 
({\it iii}) There is a GIM suppression of order $(m_c/m_b)^2$
reflecting the fact that, in the limit $m_c = m_u$, any CP asymmetry
in the SM vanishes. It will be rather difficult to make an inclusive
measurement of the CP asymmetry in the $b \to d$ channel. However,
based on CKM unitarity, one can derive a U-spin relation between the
direct CP asymmetries in the $b \to d$ and the $b \to s$
channel~\cite{Soares}. U-spin breaking effects are estimated to be
negligibly small.  This finally leads to the SM zero-prediction for
the CP asymmetry in the untagged mode $\bar B \rightarrow X_{s+d}
\gamma$~\cite{mannelhurth1,mannelhurth2}. This zero prediction
provides a very clean test, whether new CP phases are active or
not. Any significant deviation from this prediction would be a direct
hint of non-CKM contributions to CP violation.

In the present paper we analyse this relation within various general
scenarios beyond the SM. In~\cite{recksiegel} the untagged CP
asymmetry was already considered in a specific model with vector
quarks. We will first focus on supersymmetric models with minimal
flavour violation and then consider a very general parameterization of
new physics contributions. The first part of our analysis is, thus,
based on the consistent definition of minimal flavour violation
recently presented in~\cite{Giannew} in which all flavour and
CP-violating interactions originate from the Yukawa couplings. This
constraint is introduced using an effective field approach
supplemented by a symmetry concept and can be shown to be
renormalization-group invariant. We shall also extend consistently
this definition by introducing flavour-blind phases. We realize the
minimal flavour violating scenario in terms of a flavour-blind minimal
supersymmetric standard model (MSSM) in which all the soft breaking
terms are generated at the GUT scale. The case of general flavour
violation is analysed in a model-independent framework. Taking into
account the available experimental bounds on the $\bar B \rightarrow
X_s \gamma$ branching ratio and on the corresponding direct CP
asymmetry, we analyse possible correlations between the tagged and
untagged measurements and their indirect sensitivity to the asymmetry
in the $\bar B \rightarrow X_d \gamma$.

We also derive general model-independent formulae for the branching
ratios and the direct CP asymmetries for the two inclusive modes $\bar
B \rightarrow X_s \gamma$ and $\bar B \rightarrow X_d \gamma$, which
also allow us to update the corresponding SM predictions.

Let us summarize the experimental situation and prospects.  The
present experimental accuracy on the inclusive decay $\bar B
\rightarrow X_s \gamma$ already reached the $10 \%$ level, as
reflected in the world average of the present
measurements~\cite{aleph,belle,cleobsg,babar1,babar2,jessop}:
\beq
{\cal B}[\bar B \to X_s \gamma] = (3.34 \pm 0.38) \times 10^{-4} \,.
\label{world}
\eeq
In the near future, more precise data on this mode are expected from
the $B$-factories. In particular, the direct CP asymmetries are now
within experimental reach. The first measurement of this asymmetry was
presented by CLEO~\cite{cleoCP}; it is actually a weighted sum over
the $b\to s$ and $b\to d$ channels: $A_{\rm CP} = 0.965\, A_{\rm CP}[\bar B
\rightarrow X_s \gamma] + 0.02\, A_{\rm CP}[\bar B \rightarrow X_d
\gamma]$, yielding
\begin{equation}
A_{\rm CP}[\bar B \rightarrow X_s \gamma] = (-0.079 \pm 0.108 \pm 0.022) \times (1.0 \pm 0.030) \,.
\label{acpexp}
\end{equation}
The first error is statistical, the second and third errors are
additive and multiplicative systematics respectively. This measurement
is based on $10^{7}$ $B \bar B$ events (on resonance); it uses fully
inclusive and semi-inclusive techniques and implies $-0.27 <
A_{\rm CP}[\bar B \rightarrow X_s \gamma] < +0.10$ at $90\%$ confidence
level.  The recent Belle measurement~\cite{belleCP} uses
semi-inclusive techniques; it is based on $15 \times 10^{7}$ $B \bar
B$ events (on resonance) and leads to
\begin{equation}
A_{\rm CP}[\bar B \rightarrow X_s \gamma] = -0.004 \pm 0.051  \pm 0.038 \,,
\label{acpexp2}
\end{equation}
where the first error is statistical and the second systematic. This
corresponds to $-0.107 < A_{\rm CP} < 0.099$ at $90 \%$ confidence level.
Very large effects are thus already experimentally excluded. Note that
the same conclusion can be deduced from the measurements of the CP
asymmetry in the exclusive $B \rightarrow K^* \gamma$ modes. The world
average includes CLEO, Babar and Belle measurements and
reads~\cite{nakaoCP}.
\begin{equation}
A_{\rm CP}[B \rightarrow K^* \gamma] = -0.005 \pm 0.037 \, .  
\end{equation} 
We stress that the application of quark--hadron duality is, in
general, problematic within a semi-inclusive measurement of
CP-violating effects, if only 50\% or 70\% of the total exclusive
modes are detected. In fact, the strong rescattering phases
responsible for the presence of CP violation can be different for each
exclusive channel. It is impossible to reliably quantify the resulting
systematic uncertainty without a detailed study of the individual
modes and of their direct CP asymmetries. Therefore, a fully inclusive
measurement of the {\it untagged} direct CP asymmetry, the observable
on which the present paper focuses, is favoured. Such a measurement is
possible because the experimental efficiencies within the inclusive $b
\rightarrow s$ and $b \rightarrow d$ modes are expected to be equal.
Recent analyses of the future experimental accuracy~\cite{colinjim}
expect a total integrated luminosity of about $1 ab^{-1}$, by the end
of BaBar and Belle; this translates into an experimental error on the
CP asymmetries of order $3\%$~\footnote{For semi-inclusive
measurements the experimental error is expected to be even
smaller. However, in this case, the additional systematic
uncertainties discussed above must be taken into account.}.  The
potential of the so-called Super-$B$-factories with an integrated
luminosity of about $50 ab^{-1}$ would even lead to an experimental
uncertainty of about $0.5 \%$~\cite{colinjim}.

The plan of this paper is as follows: in the next section we derive in
detail the improved model-independent formulae for the branching
ratios and the direct tagged CP asymmetries for the two inclusive
modes $\bar B \rightarrow X_s \gamma$ and $\bar B \rightarrow X_d
\gamma$. Here we also present our new SM predictions there. In
Section~3 we discuss the untagged CP asymmetry within the SM. In
Section~4 we present our analysis within the minimal flavour violation
scenario. Finally, Section~5 contains the corresponding
model-independent analysis -- followed by our conclusions.

\section{$B \rightarrow X_s, X_d \gamma$ decays beyond the SM:\\
branching ratios and CP asymmetries}
In this section we derive general model-independent formulae for the
branching ratios and the direct tagged CP asymmetries for the
inclusive $\bar B \rightarrow X_{s,d} \gamma$ modes.  Our main aim is
to present explicit numerical expressions for these observables as
functions of Wilson coefficients and CKM angles. The extraction of the
latter from experimental data depends critically on the assumptions
about the presence and the structure of new physics contributions to
observables such as $\Delta M_{B_{d}}$, $\Delta M_{B_{s}}$,
$\epsilon_K$, $a_{\psi K_s}$.  Therefore, the numerical expressions
that we will present below in Eqs.~(\ref{brnum}) and (\ref{acpnum})
will be very useful in phenomenological analyses.

For this purpose we generalize the SM results at the NLL level given
in Ref. \cite{GM}~\footnote{Reference \cite{GM} presents a detailed
discussion of the NLL QCD formulae, which are based on the original
NLL QCD calculations presented in \cite{Adel,GHW,Mikolaj} and on
independent checks of these calculations \cite{GH,Burasnew,Paolonew}.}
in order to accommodate new physics models with new CP-violating
phases as well as implement several improvements.  We also update the
corresponding SM predictions.

Let us start with the generalization of the NLL formulae and also the
discussion of the input parameters:

\begin{itemize} 

\item 
The general effective hamiltonian that governs the inclusive $\bar
B\to X_q \gamma$ decays ($q=d,s$) in the SM is
\begin{equation}
H_{\rm eff} (b\rightarrow q \gamma) =
- {4 G_F\over \sqrt{2}} V_{tb}^{} V_{tq}^* \left( \sum_{i=1}^8 C_i (\mu)\cdot O_i (\mu) 
+ \epsilon_q \, \sum_{i=1}^2 C_i (\mu)\cdot (O_i (\mu)-O_i^u (\mu) )\right),
\label{bqgEH}
\end{equation}
where $\epsilon_q = (V_{ub}^{} V_{uq}^*) / (V_{tb}^{} V_{tq}^*)$ and
the operators are: \\
\noindent
\begin{minipage}{0.45\linewidth}
\begin{eqnarray}
\hskip -0.5cm && O_1^u = (\bar{q}_{L} \gamma_\mu T^a u_{ L}) (\bar{u}_{ L} \gamma^\mu T^a b_{ L}), \nn \\
\hskip -0.5cm && O_1  = (\bar{q}_{L} \gamma_\mu T^a c_{ L}) (\bar{c}_{ L} \gamma^\mu T^a b_{ L}),\nn  \\
\hskip -0.5cm && O_3 = (\bar{q}_{L} \gamma_\mu b_{ L}) \textstyle \sum_{q'} ({\bar{q}'}_{ L} \gamma^\mu {q'}_{ L}),\nn \\
\hskip -0.5cm && O_5 = (\bar{q}_{L} \gamma_\mu \gamma_\nu \gamma_\rho b_{ L}) \textstyle \sum_{q'} ({\bar{q}'}_{ L}
         \gamma^\mu \gamma^\nu \gamma^\rho {q'}_{ L}),\nn \\
\hskip -0.5cm && O_7 =\frac{e }{16 \pi^2} m_b (\mu) (\bar{q}_{L} \sigma_{\mu \nu} b_{R}) F^{\mu \nu},\nn 
\end{eqnarray}
\end{minipage}
\begin{minipage}{0.51\linewidth}
\begin{eqnarray}
\hskip -0.5cm && O_2^u = (\bar{q}_{L} \gamma_\mu u_{ L}) (\bar{u}_{ L} \gamma^\mu b_{ L}),\nn\\
\hskip -0.5cm && O_2 = (\bar{q}_{L} \gamma_\mu c_{ L}) (\bar{c}_{ L} \gamma^\mu b_{ L})\nn\\
\hskip -0.5cm && O_4 = (\bar{q}_{L} \gamma_\mu T^a b_{ L}) \textstyle \sum_{q'} ({\bar{q}'}_{ L} \gamma^\mu T^a {q'}_{ L}),\label{operatorbasis} \\
\hskip -0.5cm && O_6 = (\bar{q}_{L} \gamma_\mu \gamma_\nu \gamma_\rho T^a b_{ L}) \textstyle \sum_{q'} ({\bar{q}'}_{ L}
         \gamma^\mu \gamma^\nu \gamma^\rho T^a {q'}_{ L}),\nn\\
\hskip -0.5cm && O_8  = \frac{g_s }{16 \pi^2} m_b (\mu) (\bar{q}_{L} T^a \sigma_{\mu \nu} b_{R }) G^{a \mu \nu} \; . \nn
\end{eqnarray}
\end{minipage}

We assume, within our model-independent analysis, that the dominant
 new physics effects only modify the Wilson coefficients of the dipole
 operators $O_7$ and $O_8$ and also introduce contributions
 proportional to the corresponding operators with opposite chirality:
\begin{equation}
O_7^R = \frac{e }{16 \pi^2} m_b (\mu) (\bar{q}_{R} \sigma_{\mu \nu} b_{L}) F^{\mu \nu}\,,  \quad  \quad 
O_8^R = \frac{g_s }{16 \pi^2} m_b (\mu) (\bar{q}_{R} T^a \sigma_{\mu \nu} b_{L}) G^{a \mu \nu} \; .
\label{operatorchiral}
\end{equation}

\item 
The branching ratio for $\bar B\to X_q \gamma$ can be parameterized as
\bea
\branch [\bar B \to X_q \gamma]_{E_\g> E_0}^{{\rm subtracted} \psi,\psi'} 
&=&
\branch [\bar B \to X_c e \bar\nu]_{\rm exp} \, {6 \alpha_{\rm em} \over \pi C} \,
\left| V_{tq}^* V_{tb}^{}\over V_{cb}^{}\right|^2 
\, \Big[ P(E_0) + N(E_0) \Big] \\
& = & {\cal N}\,  \left| V_{tq}^* V_{tb}^{}\over V_{cb}^{}\right|^2  \, \Big[ P(E_0) + N(E_0) \Big] \, ,
\label{br}
\eea
where $P(E_0)$ and $N(E_0)$ denote the perturbative and the
non-perturbative contributions respectively.

\item
The pre-factor $C$ in Eq.~(\ref{br})  is given by 
\beq
C = \left| V_{ub} \over V_{cb} \right|^2 { \Gamma 
[\bar B\to X_c e \bar \nu] \over \Gamma [\bar B\to X_u e \bar \nu] } \,.
\eeq
In Ref.~\cite{GM}, the authors present a detailed determination of $C$
employing the $\Upsilon$ expansion~\cite{upsilon}. The uncertainties
on $C$ are dominated by the errors on the heavy quark effective theory
parameter $\lambda_1$ and on the non-perturbative contribution to the
$\Upsilon$ mass, $\Delta = m_{\Upsilon}/2 - m_b^{1S}$, where
$m_b^{1S}$ is defined as half of the perturbative contribution to the
$\Upsilon$ mass. Using $\lambda_1 = (-0.27 \pm 0.10 \pm 0.04) \, {\rm
GeV}^2$~\cite{upsilon} and $m_b^{1S} = (4.69 \pm 0.03 )\, \gev
$~\cite{hoang}, from which follows $\Delta^2 /(m_{\Upsilon}/2) =( 0.04
\pm 0.03 ) \, {\rm GeV}$, the authors of ref.~\cite{GM} obtain $C =
0.575 \, (1 \pm 0.01_{\rm pert} \pm 0.02_{\lambda_1} \pm
0.02_{\Delta}) = 0.575 \, (1 \pm 0.03)$. In this analysis we prefer to
increase the controversially small error on $m_b^{1s}$ given in
Ref.~\cite{hoang} (see the discussion in Sec.~7 of
Ref.~\cite{BS}). Moreover we increase the error on $\lambda_1$ in
order to include the effects of the unknown $\Lambda_{\rm
QCD}^3/m_D^2$ corrections to the $m_c^{\rm pole} / m_b^{\rm pole}$
ratio.  Our more conservative error analysis leads us to $C = 0.575 \,
(1 \pm 0.01_{\rm pert} \pm 0.04_{\lambda_1} \pm 0.04_{\Delta}) = 0.575
\, (1 \pm 0.06)$. Using $\branch [\bar B \to X_c e \bar\nu]_{\rm exp}
= 0.1074 \pm 0.0024$ and $\alpha_{\rm em} = 1/137.036$, we finally
obtain
\beq
{\cal N} = 2.567 \, (1 \pm 0.064 ) \times 10^{-3} \, .
\eeq

\item 
In this paper we use  the Wolfenstein parameters of the CKM matrix 
according to the fit presented in Ref.~\cite{yellowbook} ($\lambda =
0.2240 \pm 0.0036$, $A = 0.83 \pm 0.02$, $\bar \rho = 0.162 \pm
0.046$, $\bar \eta = 0.347 \pm 0.027$) and obtain 
\bea
\left| V_{ts}^* V_{tb}^{}\over V_{cb}^{}\right|^2 &=& 0.9648 \, ( 1 \pm 0.005 ) \, , \\
\left| V_{td}^* V_{tb}^{}\over V_{cb}^{}\right|^2 &=& 0.0412 \, ( 1 \pm 0.10  ) \, \label{normd} \\ 
V_{us}^* V_{ub}^{}\over V_{ts}^{*}  V_{tb}^{}   &=&  \epsilon_s = (-0.0088 \pm 0.0024) + i\, (0.0180 \pm 0.0015) \,, \\
 V_{ud}^* V_{ub}^{}\over V_{td}^{*}  V_{tb}^{}    &=&   \epsilon_d = (0.019 \pm 0.046) - i\, (0.422 \pm 0.046) \,.
\eea
From these values it is obvious that CKM uncertainties are completely
negligible in $b\to s \gamma$ transitions but play an important role
in $b\to d \gamma$ ones.

\item 
The non-perturbative contribution $N(E_0)$ in Eq.~(\ref{br}) is not
sensitive to new physics and we will use the numerical estimate of
Ref.~\cite{GM}:

\begin{equation}
N(E_0) = 0.0036 \pm 0.0006.
\end{equation}

\item 
The perturbative contribution $P(E_0)$ is defined by \cite{GM}
\bea
{ \Gamma [b\to X_q \gamma]_{E_\gamma > E_0} \over 
|V_{cb} / V_{ub}|^2 \, \Gamma [b\to X_u e \bar \nu]} 
= \left| V_{tq}^* V_{tb}^{} \over V_{cb} \right|^2 {6 \alpha_{\rm em} \over \pi} P(E_0) \,.
\eea
This contribution can be parameterized in the following way:
\bea
P(E_0) &=& \left| K_c + \left(1 + {\alpha_s (\mu_0) \over \pi} \log {\mu_0^2 \over m_t^2} \right) 
           r (\mu_0) K_t + \varepsilon_{\rm ew} \right|^2 + B(E_0) \,,
\label{pert}
\eea
where $K_t$ is the top contribution to the $b\to q \gamma$ amplitude,
$K_c$ contains the remaining contributions (including those from the
operators $O_i^u$) and $\varepsilon_{\rm ew}$ are the electroweak
corrections; $r (\mu_0) $ is the ratio of the $\overline{\rm MS}$
running mass of the bottom quark ($m_b^{\overline{\rm MS}} (\mu_0)$)
to the 1S mass ($m_b^{1S}$). The expression $r(\mu_0)$ can be found in
the appendix; $\mu_0$ denotes the matching scale, typically $m_W$.
Finally, the function $B(E_0)$ contains the bremsstrahlung effects
coming from the process $b\to q \gamma g$.

\item 
The contribution $K_c$ is practically insensitive to new physics
because the dominant new physics contributions change the dipole
operator contributions only. The explicit form of $K_c$ we use here is
slightly different from the one presented in Ref.~\cite{GM}; in fact,
the inclusion of all two-loop matrix elements of the 4-quark
operators, given in \cite{BCMU}, requires a small modification of the
term proportional to $\epsilon_q$:
\bea
K_c & = & K_c^{(0)} + K_c^{(11)} + i K_c^{(12)} + \left(K_c^{(13)} + i K_c^{(14)}\right)\, \epsilon_q \,,  
\label{kc0}
\eea
where the quantities $K_c^{ij}$ are real and their explicit
expressions are given in the appendix; $K_c^{(0)}$ and the sum of the
$K_c^{(1j)}$ are the LL and NLL contributions, respectively.  Only the
terms $K_c^{(13)}$ and $K_c^{(14)}$ differ from Ref.~\cite{GM}. This
modification is only relevant to the $b \rightarrow d$ mode and is
therefore not included in the NLL formulae of \cite{BCMU}.

\item  
The top contribution $K_t$, however, must be generalized to include
new physics with generic CP-violating phases. As already explained in
Ref.~\cite{GM} its expression coincides with Eq.~(5.1) of
Ref.~\cite{GM} after the substitutions:
\beq
C_{7,8}^{(0) {\rm SM}} (\mu_0) \rightarrow C_{7,8}^{(0) {\rm SM}} (\mu_0) + C_{7,8}^{(0) {\rm NP}} (\mu_0) = C_{7,8}^{(0) {\rm tot}} (\mu_0)  
\label{ruleNP}
\eeq
where
\beq
C_7^{(0) {\rm SM}} (\mu_0) = -{1\over 2} A_0^t (x_t) - {23\over 36}  \;\;\; \hbox{and} \;\;\;  
C_8^{(0) {\rm SM}} (\mu_0) = -{1\over 2} F_0^t (x_t) - {1\over 3} \, .
\eeq
with $x_t = (m_t (\mu_0)/m_W )^2$. The functions $A_0^t$ and $ F_0^t$
are explicitly defined in~\cite{GM}. Note that the terms proportional
to $\log (\mu_0/m_t)$ have to be treated carefully \cite{GM}.

It is convenient to parameterize the result as follows:
\bea
K_t & = & K_t^{(0)} + K_t^{(1)} + i K_t^{(1)i} \,,
\label{kttot}
\eea
where $K_t^{(0)}$ is the total LL  contribution and $K_t^{(1)} + i
K_t^{(1i)}$ is the NLL  one. The explicit $i$ factor is the only strong
phase present in the whole top contribution. These contributions can
be further decomposed in terms of the total Wilson coefficients of the
operators $O_{7,8}$ evaluated at the matching scale $\mu_0$:
\bea
 K_t^{(0)} &=&  K_t^{(01)} +  K_t^{(02)}  \, C_7^{(0) {\rm tot}} (\mu_0) + 
                           K_t^{(03)} \, C_8^{(0) {\rm tot}} (\mu_0)   \,,\label{kt0}\\
 K_t^{(1)} & = &   K_t^{(11)} 
                            +  K_t^{(12)} \, C_7^{(0) {\rm tot}} (\mu_0)  
                            +  K_t^{(13)}\, C_8^{(0) {\rm tot}} (\mu_0)  \,, \\
 K_t^{(1)i} & = &   K_t^{(14)} 
                            +  K_t^{(15)} \, C_8^{(0) {\rm tot}} (\mu_0) \,.
\label{kt1i}
\eea

We emphasize again that all the $ K_t^{(ij)}$ are real, and we list
them in the appendix. The Wilson coefficients in the SM are
numerically given by
\beq
C_7^{(0) {\rm SM}} (m_t) = -0.189
\;\;\; \hbox{and} \;\;\; 
C_8^{(0) {\rm SM}} (m_t) = -0.095 \, , 
\eeq
where we used $m_t^{\overline {\rm MS}} (m_t^{\overline {\rm MS}}) =
(165 \pm 5)\; \gev$.

In principle, one could add the NLL new physics contributions to the
NLL Wilson coefficients by replacing
$C_{7,8}^{(1) {\rm SM}} (\mu_0) \rightarrow C_{7,8}^{(1) {\rm SM}} (\mu_0) + 
C_{7,8}^{(1) {\rm NP}} (\mu_0)$. 
However, in the numerical analysis, we set $C_{7,8}^{(1) {\rm
NP}}(\mu_0)=0$ and, thus, effectively describe all new physics effects
by the LL Wilson coefficients $C_{7,8}^{\rm NP}$.

\item 
The electroweak corrections can be written as:
\bea
\varepsilon_{\rm ew} & = & \varepsilon_{\rm ew}^{\rm SM}  
                    + C_7^{(0) {\rm NP}}(\mu_0) \, \varepsilon_{\rm ew}^{(11)}
                   + C_8^{(0) {\rm NP}}(\mu_0) \, \varepsilon_{\rm ew}^{(12)} \,,
\label{epew}
\eea
where $\varepsilon_{\rm ew}^{\rm SM}= 0.0071$ is the SM
contribution~\cite{GM}; the formulae for $\varepsilon_{\rm ew}^{(ij)}$
can be found in the appendix.
\item 
The function $B(E_0)$ contains the bremsstrahlung effects coming from
$b\to s \gamma g$. Allowing for complex Wilson coefficients we get:
\bea
B(E_0) &=& {\alpha_s (\mu_b) \over \pi} {\rm Re} \Bigg\{ \sum_{i \leq j = 7}^8 
C_i^{(0){\rm eff}} (\mu_b) \, C_j^{(0){\rm eff}*} (\mu_b) \, 
\phi_{ij} (\delta,z)  \nn\\ 
 & & + \sum_{i \leq j = 1}^2 
C_i^{(0){\rm eff}} (\mu_b) \, C_j^{(0){\rm eff}*} (\mu_b) \,\Big[ 
|1+\epsilon_q|^2 \phi_{ij} (\delta,z) + |\epsilon_q|^2 \phi_{ij} (\delta,0) \Big] \nn\\
& & + \sum_{\stackrel{\scriptstyle i= 1,2}{j = 7,8}} 
C_i^{(0){\rm eff}} (\mu_b) \, C_j^{(0){\rm eff}*} (\mu_b) \,\Big[ 
(1+\epsilon_q) \phi_{ij} (\delta,z) -\epsilon_q \phi_{ij} (\delta,0) \Big] \Bigg\},
\label{be0}
\eea
where $\delta = 1 - 2 E_0/m_b$, $z=m^2_c/m^2_b$, and the
$\phi_{ij}(\delta,z)$ are given in the appendix. The Wilson
coefficients $C_j^{(0){\rm eff}}$ are given in Eq.~(E.9) of
Ref.~\cite{GM}. The inclusion of new physics contributions to the
Wilson coefficients $C_{7,8}^{(0) {\rm eff}}$ follows from the
substitutions Eq.~(\ref{ruleNP}). Note that in contrast to the virtual
contribution we have neglected the contribution of the QCD penguin
operators, $O_{3- 6}$, within the bremsstrahlung contribution whose
impact on the branching ratio is at the $0.1\%$
level~\cite{GM}. Moreover, there are additional terms coming from the
interference $O_{1,2}^u$--$O_{1,2}$ that have been pointed out for the
first time in Ref.~\cite{uc}. They are only relevant to the CP
asymmetry within the $b \rightarrow d$ sector. However, we neglect
them in this analysis because their contribution is below the
per--cent level.

\item 
It is straightforward to generalize our equations to include new
physics contributions to the dipole operators with opposite chirality
given in Eq.~(\ref{operatorchiral}). Note that $O_{7,8}^R$ do not
interfere with $O_{1-8}$ and $O_{1,2}^u$, hence terms linear in
$C_{7R}^{(0)} (\mu_0)$ and $C_{8R}^{(0)} (\mu_0)$ are absent. The
quadratic terms can be easily included using the following
prescription:
\bea
C^{(0){\rm tot}}_i (\mu_0) \; C^{(0){\rm tot}*}_j (\mu_0) 
& \longrightarrow & 
C^{(0){\rm tot}}_i  (\mu_0) \; C^{(0){\rm tot}*}_j  (\mu_0) \nn \\ 
 & & + \; C^{(0){\rm tot}}_{iR}  (\mu_0) \; C^{(0){\rm tot} *}_{jR}  (\mu_0) \;\;\; (i,j=7,8) \,.
\label{replacement}
\eea

\item 
Let us briefly comment on the choices of the scales $\mu_b$ and
$\mu_0$. We take $\mu_b = m_b^{1S}$ and vary it by a factor of 2.
Following Ref.~\cite{GM} we use $\mu_0 = m_W$ in $K_c$ and $B(E_0)$,
while $\mu_0 = m_t^{\overline{\rm MS}}(m_t^{\overline{\rm MS}})$ in
$K_t$ and $r(\mu_0)$. For the new physics contributions to the Wilson
coefficients we also use $\mu_0 =m_t^{\overline{\rm
MS}}(m_t^{\overline{\rm MS}})$.

\item 
The issue regarding the choice of the charm mass definition in the
matrix element of $O_2$ deserves a discussion. In Ref.~\cite{GM}, it
is argued that all the factors of $m_c$ come from propagators
corresponding to charm quarks that are off-shell by an amount $\mu^2
\sim m_b^2$. Therefore it seems more reasonable to use the
$\overline{\rm MS}$ running charm mass at a scale $\mu$ in the range
$(m_c,m_b)$. Here and in the following the reference values of the
charm and bottom masses are $m_c = m_c^{\rm \overline MS} (m_c^{\rm
\overline MS}) = (1.25\pm 0.10) \, \gev$ and $m_b = m_b^{1S}$.  We
first fix the central value of $m_c = 1.25\, \gev$ and vary $\mu$;
then we add in quadrature the error on $m_c$ ($\delta_{m_c} =
8\%$). The resulting determination is:
\beq
{m_c \over m_b} = 0.23 \pm 0.05 \, .
\label{myz}
\eeq
The pole mass choice corresponds, on the other hand, to ${m_c \over
  m_b} = 0.29 \pm 0.02$. Note that the question of whether to use the
  running or the pole mass is, strictly speaking, a NNLL issue. The
  most conservative position consists in accepting any value of
  $m_c/m_b$ that is compatible with any of these two determinations:
  $0.18 \leq m_c/m_b \leq 0.31$. Taking into account all past NLL  
  computations, we strongly believe that the central value $m_c/m_b
  =0.23$ represents the best possible choice, but we allow for a large
  asymmetric error that fully covers the above range (and that reminds
  us of this problem that can be solved only via a NNLL computation):
\beq
{m_c \over m_b} = 0.23_{-0.05}^{+0.08} \, .
\label{ourz}
\eeq
\end{itemize}
After this discussion of the necessary generalizations of the NLL
formulae, we present our SM updates of the branching ratios, the CP
asymmetries, and our formulae for the model-independent NLL analysis.
\begin{itemize}

\item 
We collect first the SM predictions for the branching ratios using
Eq.~(\ref{ourz}) and two different energy cuts: $E_0 = (1.6 \, {\rm
GeV} , {m_b / 20})$.  There are four sources of uncertainties: the
charm mass ($\delta_{m_c/m_b}$), the CKM factors ($\delta_{\rm CKM}(s)
= 0.5\%$, $\delta_{\rm CKM}(d) = 11\%$), the parametric uncertainty,
including the one due to the overall normalization ${\cal N}$,
$\alpha_s$ and $m_t$ ($\delta_{\rm param.}$) and the perturbative
scale uncertainty ($\delta_{\rm scale}$). Concerning the latter two
errors, we follow the analysis of Ref.~\cite{GM} and use $\delta_{\rm
param.} = 6.4\%$ and $\delta_{\rm scale}=4\%$. \\
\noindent For $E_\gamma > 1.6 \, {\rm GeV}$ we get:
\bea 
\branch [\bar B \to X_s \gamma]
& = & \left( 3.61 \,\, \left. {}^{+0.24}_{-0.40} \right|_{m_c \over m_b}
                     \pm 0.02_{\rm CKM} \pm 0.24_{\rm param.} \pm 0.14_{\rm scale} \right) \times 10^{-4}  \,, \\
\branch [\bar B \to X_d \gamma]
& = & \left( 1.38 \,\,  \left. {}^{+0.14}_{-0.21}   \right|_{m_c \over m_b}
                     \pm 0.15_{\rm CKM} \pm 0.09_{\rm param.} \pm 0.05_{\rm scale} \right) \times 10^{-5}  \,, \\
{\branch [\bar B \to X_d \gamma] \over \branch [\bar B \to X_s \gamma]}
& = & \left( 3.82 \,\,  \left. {}^{+0.11}_{-0.18}   \right|_{m_c \over m_b}
                     \pm 0.42_{\rm CKM} \pm 0.08_{\rm param.} \pm 0.15_{\rm scale} \right) \times 10^{-2}  \,.
\eea
For $E_\gamma > m_b/20$ we get:
\bea 
\branch [\bar B \to X_s \gamma]
& = & \left( 3.79 \,\,  \left. {}^{+0.26}_{-0.44}   \right|_{m_c \over m_b}
                     \pm 0.02_{\rm CKM} \pm 0.25_{\rm param.} \pm 0.15_{\rm scale} \right) \times 10^{-4}  \, , \\
\branch [\bar B \to X_d \gamma]
& = & \left( 1.46 \,\,  \left. {}^{+0.15}_{-0.23}   \right|_{m_c \over m_b}
                     \pm 0.16_{\rm CKM} \pm 0.10_{\rm param.} \pm 0.06_{\rm scale} \right) \times 10^{-5}  \, , \\
{\branch [\bar B \to X_d \gamma] \over \branch [\bar B \to X_s \gamma]}
& = & \left( 3.86 \,\,  \left. {}^{+0.11}_{-0.18}   \right|_{m_c \over m_b}
                     \pm 0.43_{\rm CKM} \pm 0.09_{\rm param.} \pm 0.15_{\rm scale} \right) \times 10^{-2}  \,.
\eea
Note that the errors on the ratio $R_{ds}=\branch [\bar B \to X_d
\gamma] / \branch [\bar B \to X_s \gamma]$ are dominated by CKM
uncertainties. We remind the reader that, on top of the mentioned
sources of error, the $B\to X_d \gamma$ mode is affected by the
presence of non-perturbative $u$-quark loops whose effect is expected
to be at most around $10\%$ (see section VI.B of Ref.~\cite{tobias}
for a more detailed discussion).
\item 
The direct CP asymmetries in $\bar B \to X_q \gamma$ can immediately
be extracted from the explicit expression Eq.~(\ref{br}) for the
branching ratio:
\bea
A_{\rm CP}^{b\to q \gamma} & \equiv & { \Gamma{[\bar B \to X_{q} \gamma]}
   - \Gamma{[B \to X_{\bar q} \gamma]} \over
      \Gamma{[\bar B \to X_q \gamma]} + \Gamma{[B \to X_{\bar q} \gamma]}} \\ 
& = & {\Delta \Gamma_{q\gamma} + \Delta \Gamma_{q g \gamma} \over  P(E_0) },
\label{acp1}
\eea
where $P(E_0)$ is defined in Eq.~(\ref{pert}) and
$\Delta\Gamma_{q\gamma,q g \gamma}$ are the contributions
corresponding to the two terms in $P(E_0)$ (virtual corrections and
bremsstrahlung) and are given in the appendix.  The rationale for
normalizing the CP asymmetry, using the complete NLL  expression for
the CP-averaged branching ratio, relies on the observation that
$\Gamma{[\bar B \to X_q \gamma]}$ and $\Gamma{[B \to X_q \gamma]}$ are
distinct observables: we are, therefore, allowed to compute them
independently to the best of our knowledge.

The SM predictions are essentially independent of the photon energy
cut-off ($E_0$) and read (for $E_0=1.6 \, {\rm GeV}$):
\bea 
A_{\rm CP} [\bar B \to X_s \gamma]
& = & 
\left( 0.44 \,\,\left. {}^{+0.15}_{-0.10} \right|_{m_c \over m_b}\, 
   \pm 0.03_{\rm CKM}   \left. {}^{+0.19}_{-0.09} \right|_{\rm scale} \right)
 \% \,, \\ 
A_{\rm CP} [\bar B \to X_d \gamma] 
& = & 
\left( -10.2 \,\,\left. {}^{+2.4}_{-3.7} \right|_{m_c \over m_b}\, 
  \pm 1.0_{\rm CKM} \left. {}^{+2.1}_{-4.4} \right|_{\rm scale} \right) \% \,.
\eea
The additional parametric uncertainties are subdominant.

\item 
Finally, we present our formulae for the branching ratios and CP
asymmetries, in which the Wilson coefficients $C_{7,8(R)}$ and all the
CKM ratios are left unspecified:
\bea
\branch [\bar B \to X_q \gamma]
& = & 
{{\cal N}\over 100} \,  \left| V_{tq}^* V_{tb}^{}\over V_{cb}^{}\right|^2  \, \Big[
a  + a_{77} \, (|R_7|^2 + |\widetilde R_7|^2)+ a_7^r \, {\rm Re} (R_7) + a_7^i \, {\rm Im} (R_7) \nonumber \\ 
& & \hskip -1.9cm 
+ a_{88} \, (|R_8|^2+ |\widetilde R_8|^2) + a_8^r \, {\rm Re} (R_8) + a_8^i \, {\rm Im} (R_8)
+ a_{\epsilon\epsilon} \, |\epsilon_q|^2 + a_\epsilon^r \, {\rm Re} (\epsilon_q) + a_\epsilon^i \, {\rm Im} (\epsilon_q) 
\nonumber \\
 & &  \hskip -1.9cm 
+ a_{87}^r \, {\rm Re} (R_8^{} R_7^* + \widetilde R_8^{} \widetilde R_7^*) + a_{7\epsilon}^r \, {\rm Re} (R_7^{} \epsilon_q^*) + 
      a_{8\epsilon}^r\, {\rm Re} (R_8^{} \epsilon_q^*) 
\nonumber \\
& &  \hskip -1.9cm 
+ a_{87}^i \, {\rm Im} (R_8^{} R_7^*+ \widetilde R_8^{} \widetilde R_7^*) 
+ a_{7\epsilon}^i \, {\rm Im} (R_7^{} \epsilon_q^*) + a_{8\epsilon}^i \, {\rm Im} (R_8^{} \epsilon_q^*) 
\Big] \, ,
\label{brnum}
\eea
\bea
A_{\rm CP}^{b\to q \gamma}
& = & \frac{{\rm Im} \Big[
a_7^i \, R_7 + a_8^i \, R_8 + a_\epsilon^i \, \epsilon_q 
+ a_{87}^i \, (R_8^{} R_7^* + \widetilde R_8^{} \widetilde R_7^*)
+ a_{7\epsilon}^i \, R_7^{} \epsilon_q^* + a_{8\epsilon}^i \, R_8^{} \epsilon_q^* \Big]}{
\displaystyle {100\over {\cal N}} \,  \left| V_{cb}^{} \over V_{tq}^* V_{tb}^{} \right|^2 
{1\over 2} \left( 
\branch [\bar B \to X_q \gamma] +
\branch [B \to X_q \gamma] 
\right)
}\,, \nn \\
\label{acpnum}
\eea
where
\bea
R_{7,8} = {C_{7,8}^{(0) {\rm tot}}(\mu_0) \over C_{7,8}^{(0) {\rm SM}}(m_t)}  
\quad \hbox{and} \quad
\widetilde R_{7,8} = {C_{7R,8R}^{(0) {\rm NP}}(\mu_0) \over C_{7,8}^{(0) {\rm SM}}(m_t)} 
\eea
and the CP conjugate branching ratio, $\branch [B \to X_q \gamma]$,
can be obtained by Eq.~(\ref{brnum}) by replacing ${\rm Im} (...)
\rightarrow -{\rm Im} (...) $. Explicit expressions for these
coefficients in terms of the quantities introduced in
Eqs.~(\ref{kc0}),(\ref{kt0})-(\ref{kt1i}),(\ref{epew}) can now be
easily derived.  Their numerical values are collected in
Table~\ref{tab:coeff}.
\end{itemize}

\begin{table}
\begin{center}
\begin{tabular}{|c|cc|cc|c|} \hline
\spa  & \multicolumn{4}{c|}{NLL} & LL \cr \hline
\spa $E_0  $ & \multicolumn{2}{c|}{$1.6 \, \hbox{GeV}$} & \multicolumn{2}{c|}{$m_b/20$} &- \cr
\spa $m_c/m_b  $ & 0.23 &0.29 & 0.23 & 0.29 & - \cr \hline\hline
\spa $ a $ & 7.8221& 6.9120& 8.1819& 7.1714& 7.9699\cr 
\spa $ a_{77}$& 0.8161& 0.8161& 0.8283& 0.8283& 0.9338\cr 
\spa $ a_7^r $& 4.8802& 4.5689& 4.9228& 4.6035& 5.3314\cr 
\spa $ a_7^i $& 0.3546& 0.2167& 0.3322& 0.2029& 0\cr 
\spa $ a_{88}$& 0.0197& 0.0197& 0.0986& 0.0986& 0.0066\cr 
\spa $ a_8^r $& 0.5680& 0.5463& 0.7810& 0.7600& 0.4498\cr 
\spa $ a_8^i $&-0.0987&-0.1105&-0.0963&-0.1091& 0\cr 
\spa $ a_{\epsilon\epsilon}$& 0.4384& 0.3787& 0.8598& 0.7097& 0\cr 
\spa $ a_\epsilon^r $&-1.6981&-2.6679&-1.3329&-2.4935& 0\cr 
\spa $ a_\epsilon^i $& 2.4997& 2.8956& 2.5274& 2.9127& 0\cr 
\spa $ a_{87}^r $& 0.1923& 0.1923& 0.2025& 0.2025& 0.1576\cr 
\spa $ a_{87}^i $&-0.0487&-0.0487&-0.0487&-0.0487& 0\cr 
\spa $ a_{7\epsilon}^r$&-0.7827&-1.0940&-0.8092&-1.1285& 0\cr 
\spa $ a_{7\epsilon}^i$&-0.9067&-1.0447&-0.9291&-1.0585& 0\cr 
\spa $ a_{8\epsilon}^r$&-0.0601&-0.0819&-0.0573&-0.0783& 0\cr 
\spa $ a_{8\epsilon}^i$&-0.0661&-0.0779&-0.0637&-0.0765& 0\cr 
\hline
\end{tabular}
\end{center}
\caption{Numerical values of the coefficients introduced in
Eqs.~(\ref{brnum}) and (\ref{acpnum}).  We give the values
corresponding to $E_0 = (1.6 \, {\rm GeV}, m_b/20)$ and to $m_c/m_b =
(0.23, 0.29)$. In the last column we give the values obtained at LL.}
\label{tab:coeff}
\end{table}

\section{Untagged $B \to X_{s+d} \gamma$ CP asymmetry}

The unnormalized CP asymmetry for the sum of the partonic processes $b
\to (s+d) \gamma$ vanishes in the limit of $m_d = m_s = 0$ as was
first observed in Ref.~\cite{Soares}. This is still valid for the
weaker condition $m_d=m_s$, which corresponds to the so-called U-spin
limit. However, any CP violation in the SM has to be proportional to
the determinant
\begin{eqnarray} \label{Jdet}
 {\rm det} \left[ {\cal M}_U \, {\cal M}_D \right]
 =  i \, J \,\,\, (m_u - m_c) (m_u - m_t) (m_c - m_t)
(m_d - m_s) (m_d - m_b) (m_s - m_b), 
\end{eqnarray}
where ${\cal M}_{U/D}$ are the mass matrices for the up and down
quarks and
\begin{equation} \label{Jarls}
J = {\rm Im}[V_{ub} V_{cb}^* V_{cs} V_{us}^*]
\end{equation}
is the Jarlskog parameter, which is the only fourth-order quantity in
the SM invariant under rephasing of the quarks fields.  If the down
and the strange quark were degenerate, the SM would be completely
CP-conserving, as can be seen from (\ref{Jdet}). Thus, the U-spin
limit at the quark level does not make much sense with respect to CP
asymmetries.  However, one shall make use of this symmetry only with
respect to the influence of the strong interactions on the hadronic
matrix elements (in particular on the strong phases), while the down
and strange quark masses are different.  The unitarity of the CKM
matrix implies
\begin{equation}
J = {\rm Im} (\lambda_u^{(s)} \lambda_c^{(s)*})
= - {\rm Im} (\lambda_u^{(d)} \lambda_c^{(d)*})\, , 
\end{equation}
where $\lambda_q^{(q')} = V_{qb}^{} V_{qq'}^{*}$. As a consequence one
finds in the U-spin limit for the hadronic matrix elements and for
real Wilson coefficients the following relation for the rate
asymmetries:
\begin{equation} \label{resexc}
\Delta \Gamma (\bar{B} \to X_s \gamma) +
\Delta \Gamma (\bar{B} \to X_d  \gamma) =  \Delta \G_s + \Delta \G_d = 0 \,, 
\end{equation}
where $\D \G_q = \D \Gamma (\bar{B} \to X_q \gamma) = \G (\bar B\to
X_q \gamma) - \G (B\to X_{\bar q} \gamma)$.

U-spin breaking effects can estimated within the heavy mass expansion
even beyond the partonic level \cite{mannelhurth1,mannelhurth2}:
\begin{equation} \label{uspinbreaking}
\Delta \Gamma (\bar{B} \to X_s \gamma) +
\Delta \Gamma (\bar{B} \to X_d  \gamma) = b_{\rm inc}\, \Delta_{\rm inc} 
\end{equation}
where the right-hand side is written as a product of a `relative
U-spin breaking' $b_{\rm inc}$ and a `typical size' $\Delta_{\rm inc}$
of the CP violating rate difference. In this framework one relies on
parton-hadron duality and one can compute the breaking of U-Spin by
keeping a non-vanishing strange quark mass.  A rough estimate of
$b_{\rm inc}$ gives a value of the order of $|b_{\rm inc}| \sim
m_s^2/m_b^2 \sim 5 \times 10^{-4}$, while $|\Delta_{\rm inc}|$ is the
average of the moduli of the two CP rate asymmetries. Thus, one
arrives at the following estimate within the partonic contribution
\cite{mannelhurth1}:
\begin{equation} \label{resinc3}
| \Delta {\cal B}(B \to X_s \gamma) +
\Delta {\cal B}(B \to X_d \gamma) | \sim 1 \cdot 10^{-9}\, .  
\end{equation}

Going beyond the leading partonic contribution within the heavy mass
expansion, one has to check if the large suppression factor from the
U-spin breaking, $b_{inc}$, is still effective in addition to the
natural suppression factors already present in the higher order terms
of the heavy mass expansion \cite{mannelhurth2}.  In the leading
$1/m_b^2$ corrections, the U-spin-breaking effects also induce an
additional overall factor $m_s^2/m_b^2$.  In the non-perturbative
corrections from the charm quark loop, which scale with $1/m_c^2$, one
finds again the same overall suppression factor, because the effective
operators involved do not contain any information on the strange mass.
Also the corresponding long-distance contributions from up-quark
loops, which scale with $\Lambda_{\rm QCD}/m_b$, follow the same
pattern \cite{tobias}.

Thus, in the inclusive mode, the right-hand side in (\ref{resinc3})
can be computed in a model-independent way, with the help of the heavy
mass expansion, and the U-spin breaking effects can be estimated to be
practically zero~\footnote{The analogous SM test within exclusive
modes is rather limited, because U-spin breaking effects cannot be
calculated in a model-independent way. Estimates
\cite{buchallabosch,mannelhurth2} lead to the conclusion that the
U-spin breaking effects are possibly as large as the rate differences
themselves.}.  Therefore, the prediction (\ref{resinc3}) provides a
very clean SM test, whether generic new CP phases are active or not.
Any significant deviation from the estimate (\ref{resinc3}) would be a
direct hint to non-CKM contributions to CP violation. This implies
that any measurement of a non-zero untagged CP asymmetry is a direct
signal for new physics beyond the SM.

A simple expression of the untagged CP asymmetry is given by:
\bea
A_{\rm CP} (B \to X_{s+d} \gamma) &=& 
{\D \G_s + \D \G_d \over \S \G_s + \S \G_d} 
= {A_{\rm CP} (B \to X_s \gamma) + R_{ds} \; A_{\rm CP} (B \to X_d \gamma)
     \over 1 + R_{ds}}
\label{untag}
\eea
where $\S \G_q = \G (\bar B\to X_q \gamma) + \G (B\to X_q \gamma)$ and
$R_{ds} = \S\G_d / \S\G_s$\, ($\D \G_q$ is defined above).

From Eq.~(\ref{untag}) again one easily derives that in the SM the
untagged asymmetry $A_{\rm CP} (B\to X_{s+d}\gamma)$ vanishes
identically: CKM unitarity and the reality of the Wilson coefficients
in the SM imply $\Delta \Gamma_s = c_s \, |V_{ts}|\, {\rm Im}
\epsilon_s $ and $\Delta \Gamma_d = c_d\, |V_{td}|\, {\rm Im}
\epsilon_d$ where $\epsilon_q = (\ds V_{uq}^{*} V_{ub}^{}) / (
V_{tq}^{*} V_{tb}^{})$. The U-spin symmetry for the hadronic matrix
elements then implies $c_s = c_d$ and one gets finally:
\begin{equation} 
A_{\rm CP}^{\rm SM} (B \to X_{s/d} \gamma) 
\propto  {\rm Im} \left( 
\epsilon_s + \left|  V_{td}^{} /   V_{ts}^{} \right|^2 \epsilon_d \right) = 0\; .
\end{equation}

We first note that recent experimental results from
BABAR~\cite{babar:convery}, put the following upper limit on the ratio
of exclusive $B$ decays $B \rightarrow K^* \gamma$ and $B \rightarrow
\rho \gamma$
\beq
R(\r\g/K^* \g) = { \G (B\to \r \g) \over \G (B\to K^* \g)} 
\leq 0.047 \hbox{ at } \cl{90}\; .
\label{ratioexclusive}
\eeq
Assuming $R_{ds} \sim R(\r\g/K^* \g)$ one can conclude that all new
physics models, in which the $B\to X_s \g$ CP asymmetry is $\sim 5\%$,
also predict a sizeable untagged asymmetry $A_{\rm CP} (B \to X_{s+d}
\gamma) \sim A_{\rm CP} (B \to X_s \gamma)$.  The only exception is
the case of a cancellation between the two terms in (\ref{untag}),
which is only possible for $A_{\rm CP} (B \to X_d \gamma) \sim 100\%$.
These considerations lead to the following general questions: (i) To
which extent are the untagged CP asymmetry $A_{\rm CP} (B \to X_{s+d}
\gamma)$ and the tagged CP asymmetry $A_{\rm CP} (B \to X_{s} \gamma)$
correlated?  (ii) To which extent can the untagged CP asymmetry
$A_{\rm CP} (B \to X_{s+d} \gamma)$ be sensitive to the CP asymmetry
in the $d$ sector $A_{\rm CP} (B \to X_{d} \gamma)$?

Clearly, predictions for the normalized CP asymmetries in $\bar B
\rightarrow X_{s/d} \gamma$ beyond the SM are rather model-dependent
\cite{summarys,summaryd}.  For example, supersymmetric predictions
depend strongly on the assumptions for the supersymmetry-breaking
sector \cite{summarys,summaryd}. However, especially for the untagged
CP asymmetry, specific properties can be identified within general
classes of models.
In the following sections we analyze the above questions in various
supersymmetric scenarios, namely in so-called minimal flavour
violation models with and without additional sources of CP violation.
Moreover, we study also general flavour violation models using a
model-independent approach.

\section{CP asymmetries within minimal flavour violation}
\label{sec:mfv}

\subsection{RG-invariant definition of MFV}

In the analysis of FCNC processes beyond the SM, especially within
supersymmetry, the additional assumption of minimal flavour violation
(MFV) is often introduced.  MFV is then loosely defined as: `all
flavour changing interactions are completely determined by the CKM
angles'.  Especially in a renormalization-group equation (RGE)
approach, the naive assumption of MFV is problematic, since it is not
stable under radiative corrections and calls for a more precise
concept. In ref.~\cite{Giannew}, a consistent definition was
presented, which essentially also requires that all flavour and
CP-violating interactions are linked to the known structure of Yukawa
couplings.  The constraint is introduced with the help of a symmetry
concept and can be shown to be RGE invariant, which is a crucial
ingredient for a consistent effective field theory approach.

In fact, it is well known that the maximal flavour symmetry group of
unitary field transformations allowed by the gauge part of the SM
Lagrangian, $U(3)^5$, can be decomposed in the following way
\begin{equation}
G_F \equiv {SU}(3)^3_q \otimes  {SU}(3)^2_\ell
\otimes  {U}(1)_B \otimes {U}(1)_L \otimes {U}(1)_Y 
\otimes {U}(1)_{PQ} \otimes {U}(1)_{E_R}~,
\end{equation}
where
\bea
{SU}(3)^3_q     &=& {SU}(3)_{Q_L}\otimes {SU}(3)_{U_R} \otimes
 {SU}(3)_{D_R}~,  \\ 
{SU}(3)^2_\ell  &=&  {SU}(3)_{L_L} \otimes {SU}(3)_{E_R}~.
\eea

The subgroup ${SU}(3)^3_q \otimes {SU}(3)^2_\ell \otimes {U}(1)_{PG}
\otimes {U}(1)_{E_R}$ is broken by the Yukawa part of the
SM. Nevertheless, one can formally promote the group $G_F$ to an exact
symmetry by assuming that the Yukawa matrices are vacuum expectation
values of dimensionless auxiliary fields $Y_U$, $Y_D$, and $Y_E$
transforming under ${\rm SU}(3)^3_q \otimes {\rm SU}(3)^2_\ell$ as
\beq Y_U \sim (3, \bar 3,1)_{{\rm SU}(3)^3_q}~,\qquad Y_D \sim (3, 1,
\bar 3)_{{\rm SU}(3)^3_q}~,\qquad Y_E \sim (3, \bar 3)_{{\rm
SU}(3)^2_\ell}~.  \eeq
By definition, an effective theory satisfies the MFV criterion if all
 higher-dimensional operators, constructed from SM and $Y$ fields, are
 invariant under CP and (formally) under the flavour group $G_F$.
 Thus, MFV requires the dynamics of flavour violation to be completely
 determined by the structure of the ordinary Yukawa couplings. This
 also means that all CP violation originates from the CKM phase
 \cite{Giannew}.
We note here that one can extend this consistent concept of MFV by
adding flavour-blind phases. In this case CP is not only broken by the
CKM phase but also by these additional phases.  However, the important
property of renormalization-group invariance of the concept of MFV is
untouched.

The hierarchical structure of the CKM matrix and of the Yukawa
couplings restricts the number of relevant operators
significantly. This leads to one of the key predictions of the MFV:
the existence of a direct link between the $b \rightarrow s$,
$b\rightarrow d$ and $s \rightarrow d$ transitions. This prediction,
within the $\Delta F = 1$ sector, is definitely not well tested at the
moment.

\subsection{Flavour-blind supersymmetric models \\
            with and without extra phases}

In supersymmetric theories a necessary condition for the fulfillment
of the MFV requirement is that all soft SUSY-breaking terms can be
diagonalized by superfield rotations. Note that this is a non-trivial
statement, because the two matrices $\tilde V_L$ and $\tilde V_R$,
which diagonalize the soft SUSY-breaking masses squared for the left
and right squarks, respectively, must also diagonalize the left--right
mixing matrix.  Provided that the above statement is correct, one can
put all the information on flavour changing couplings inside the
Yukawas of the superpotential
\begin{equation} 
Y_U^{\rm diag} = V_{CKM} D_L \; Y_U \; U_R^\dagger\, , \quad
Y_D^{\rm diag} = D_L \; Y_D \; D_R^\dagger \; .
\end{equation}
A sufficient condition for the MFV concept to be realized is that
$D_L$, $U_R$ and $D_R$ are unit matrices. If we allow for additional
phases, these are unit matrices times a phase.

We realize both options for MFV models, with and without additional
 phases, by a flavour-blind Minimal Supersymmetric standard model
 (MSSM) with conserved R-parity, where all the soft breaking terms are
 generated at the GUT scale and evolved to the electroweak scale by
 two-loop RGEs \cite{Martin:1993zk,Yamada:1994id,Jack:1994kd}.
We define the soft breaking terms at the GUT scale as
\begin{eqnarray}
\label{soft}
&& (M_a^2)_{i j} = M_A^2\ \delta_{i j} \;\;\;\; 
(a=Q,\; U,\; D,\; L,\; E) \nn\\
&& (Y^A_a)_{i j}= A_a e^{i \phi_{A_a}} (Y_a)_{i j} 
\;\;\;\; (a=U,\; D,\; E) \nn\\
&& M_{H_1}^2 ,\  M_{H_2}^2\ \nn \\
&& B e^{i \phi_B} \nn \\
&& e^{i \phi_a} M_a \;\;\;\;  (a=1,\; 2,\; 3)
\end{eqnarray} 
where $i,j$ are family indices, the $Y^A_f$ are trilinear scalar
couplings and $Y_f$ denote the Yukawa matrices; $M_a^2$ are the soft
SUSY breaking masses for the sfermions.  In contrast to the analysis
in \cite{blindpaper}, we do our analysis within the most general
flavour blind analysis and do not assume any additional constraint on
the soft breaking terms such as universality or $SU(5)$ symmetry;
$M_{H_1}$ and $M_{H_2}$ represent the Higgs soft breaking masses and
$B$ mixes both Higgs doublets.
Beside the parameters $A_a$, $B$, $M_a$ of Eq.~(\ref{soft}) also $\mu$
can be complex, yielding a total of six phases. Two of these phases
can be eliminated because of a Peccei--Quinn symmetry and an
R-symmetry \cite{Dugan:1984qf}. We work in a basis where $B$ and $M_2$
are real. For simplicity we also assume that the remaining gaugino
phases are real.

Our flavour-blind assumptions at the GUT scale (\ref{soft}) are
compatible with the general MFV scenario. In fact, the two properties,
namely that the soft contributions of the scalar mass are universal in
generation space and that the trilinear soft terms are proportional to
Yukawa couplings, are sufficient conditions for MFV.  At an arbitrary
scale, however, the physical squark masses are not equal, but the
induced flavour violation is still described in terms of the usual CKM
parameters. Having used the RG equation we arrive at the following
mass terms for left sfermions:
\begin{equation}
(M_Q^2)_{ij} = M_{Q,0}^2 \times [ \alpha^Q_0 \delta_{ij}
                            + \alpha^Q_1 (Y_U Y_U^\dagger)_{ij}
                            + \alpha^Q_2 (Y_D Y_D^\dagger)_{ij}
                            + \dots ]
\label{massterms}
\end{equation}
Owing to the hierarchical structure of the flavour parameters, the
higher order terms in $Y_a$ are numerically strongly suppressed.
Analogous statements about the other sfermion mass parameters as well
as the trilinear couplings are also valid.

As mentioned above we now consider two options in our analysis. In the
 first scenario, we put all flavour-blind phases in (\ref{soft}) to
 zero; thus, we are then in the strict MFV scenario, where the only
 source of CP violation is the CKM matrix.  In the second scenario, we
 keep the phases for the $A$-parameters and for $\mu$. However, then
 we have to take into account the constraints of the electric dipole
 moments (EDM) of the electron and of the neutron.  Contrary to the
 SM, where the EDMs occur at the higher loop only and the theoretical
 predictions are very small, the SUSY contributions appear already at
 one-loop order leading to theoretical predictions, which in general
 exceed the experimental bounds. The resulting strong constraints on
 the complex phases within SUSY models reflect the well-known SUSY-CP
 problem\footnote{Note, that the constraints for the electron EDM are
 less severe once additional phases for flavour violating parameters
 are taken into account \cite{Bartl:2003ju}.}. We will investigate how
 the EDM constraints restricts the ranges for the CP asymmetries in
 the $b$ system.

\subsection{Numerical analysis}

In this section we present our numerical results for the CP
asymmetries for the two scenarios discussed before.
Before proceeding we briefly summarize the main procedure to calculate
the parameters at the electroweak scale.  The gauge couplings $g_1$,
$g_2$, $g_3$ and the Yukawa couplings are calculated in the
$\overline{DR}$ scheme by adopting the shifts given in
\cite{bagger}. In case of the top and bottom Yukawa couplings, we
include the two-loop gluonic part \cite{Avdeev:1997sz} in the shifts.
In the case of the bottom and tau Yukawa couplings, we resum the SUSY
contributions as proposed in
\cite{Eberl:1999he,Carena:1999py,Buras:2002vd}.

These parameters are evolved to $M_{\rm GUT}$ using two-loop RGEs
\cite{Martin:1993zk,Yamada:1994id,Jack:1994kd}.  At two-loop order the
gauge couplings do not meet exactly
\cite{Weinberg:1980wa,Hall:1980kf}; the differences are due to
threshold effects at the unification scale $M_{\rm GUT}$ and leave us
with an ambiguity in the definition of $M_{\rm GUT}$. In this paper we
define $M_{\rm GUT}$ as the scale where $g_1 = g_2$ in the RGE
evolution.
At the scale $M_{\rm GUT}$ the boundary conditions for the soft
SUSY-breaking parameters are imposed. All parameters are evolved to
$M_{\rm SUSY}\equiv \sqrt{m_{\tilde t_1} m_{\tilde t_2}}$ using
two-loop RGEs. At this scale the masses of all SUSY particles are
calculated using one-loop formulae (which are a three-generation
extension of those presented in \cite{bagger}).  In the case of the
masses of the neutral Higgs bosons two-loop contributions as given in
\cite{Degrassi:2001yf,Dedes:2002dy} are included.  The absolute value
of $\mu$ is, as usual, obtained from radiative electroweak symmetry
breaking while its phase is completely arbitrary. Here we have
included the complete one-loop contributions \cite{bagger} as well as
the leading two-loop contributions given in \cite{Dedes:2002dy}. The
complete procedure is iterated until the resulting masses change by
less than one per-mill between two iterations.
For further technical details on the procedure see
\cite{Porod:2003um}.

Once a stable solution has been found, the couplings are evolved from
$M_{\rm SUSY}$ to $m_t(m_t)$, where the contributions to the Wilson
coefficients are calculated and then to $m_W$ where the EDMS are
calculated.
In the second scenario with complex phases, we take into account the
constraints due to the bounds on the neutron EDM and the electron EDM:
\begin{equation}
          |d_n|_{\rm exp} \le  6 \times 10^{-26} e \, cm\,,
  \qquad  |d_e|_{\rm exp} \le  7 \times 10^{-28} e \, cm\,.  \label{EDM}
\end{equation}
 The calculation of the neutron EDM is performed in two different
neutron models, the Chiral Quark model and the Quark--Parton model, to
get an estimate of the involved theoretical uncertainty. Here we have
used the formulae for both models as presented in
\cite{hep-ph/9903402}.
We select the points within the supersymmetric parameter space, which
are compatible with the constraint on the neutron EDM in the following
way: the phases in the trilinear terms, $\phi_{A_U}$ and $\phi_{A_D}$,
(see Eq.~(\ref{soft})) are randomly chosen, and then the phase of the
parameter $\mu$ is chosen such that the experimental constraint on the
neutron EDM is fulfilled for at least one of the two neutron
models. Finally, the experimental bound on the electron EDM, induced
by the phase of $\mu$ and $\phi_{A_E}$ can be fulfilled by an
appropriate choice of the latter phase.

For the numerical results presented below we have varied the
parameters in the following ranges:
\bea
\tan \beta & \in & [2,50] \\
M_{1/2} & \in & [ 100, 1000] \; {\rm GeV} \\
M_{H_i}, \;  M_a & \in & [ 100, 1000] \; {\rm GeV} \;\;\;\;
                  (a=Q,U,D,L,E, \; i=1,2) \\
|A_u|& \leq &  \sqrt{3(M^2_Q+ M^2_U+ M^2_{H_2}) } \\ 
|A_d|& \leq &  \sqrt{3(M^2_Q+ M^2_D+ M^2_{H_1}) } \\ 
|A_e|& \leq &  \sqrt{3(M^2_L+ M^2_E+ M^2_{H_1}) } 
\eea
The range of the $A$ parameters is restricted to avoid the danger of
colour and/or charge breaking minima.

\begin{figure}
\begin{center}
\epsfig{figure=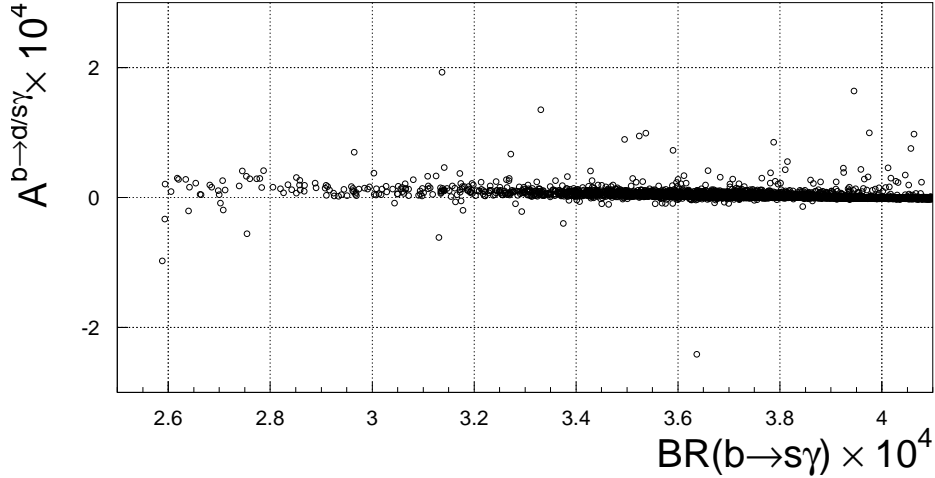,width=.8\linewidth}
\end{center}
\caption{Untagged CP asymmetry in the MFV scenario without
flavour-blind phases.}
\label{real}
\end{figure}

\begin{figure}
\begin{center}
\epsfig{figure=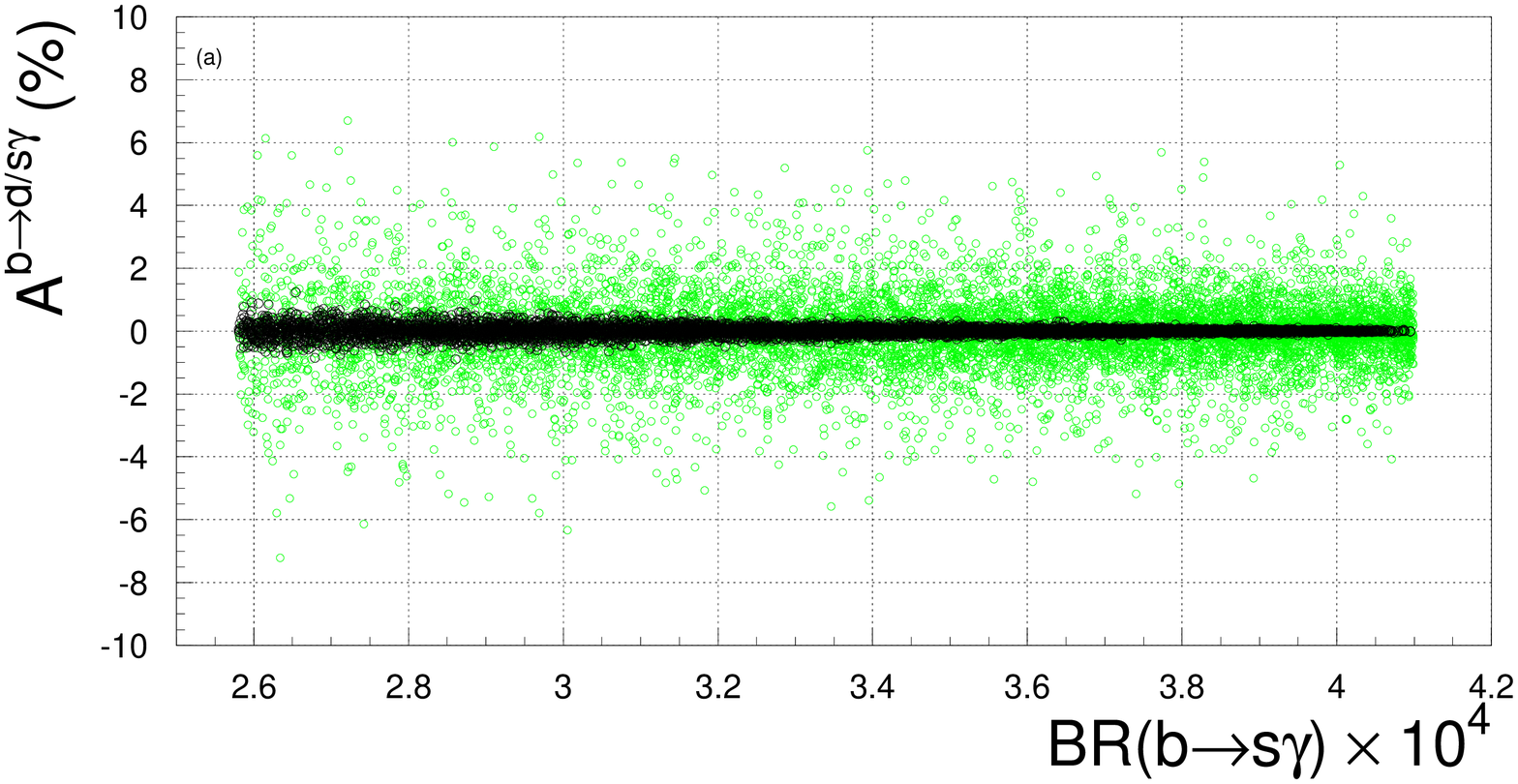,width=.8\linewidth}
\end{center}
\vspace{-1.8cm}
\begin{center}
\epsfig{figure=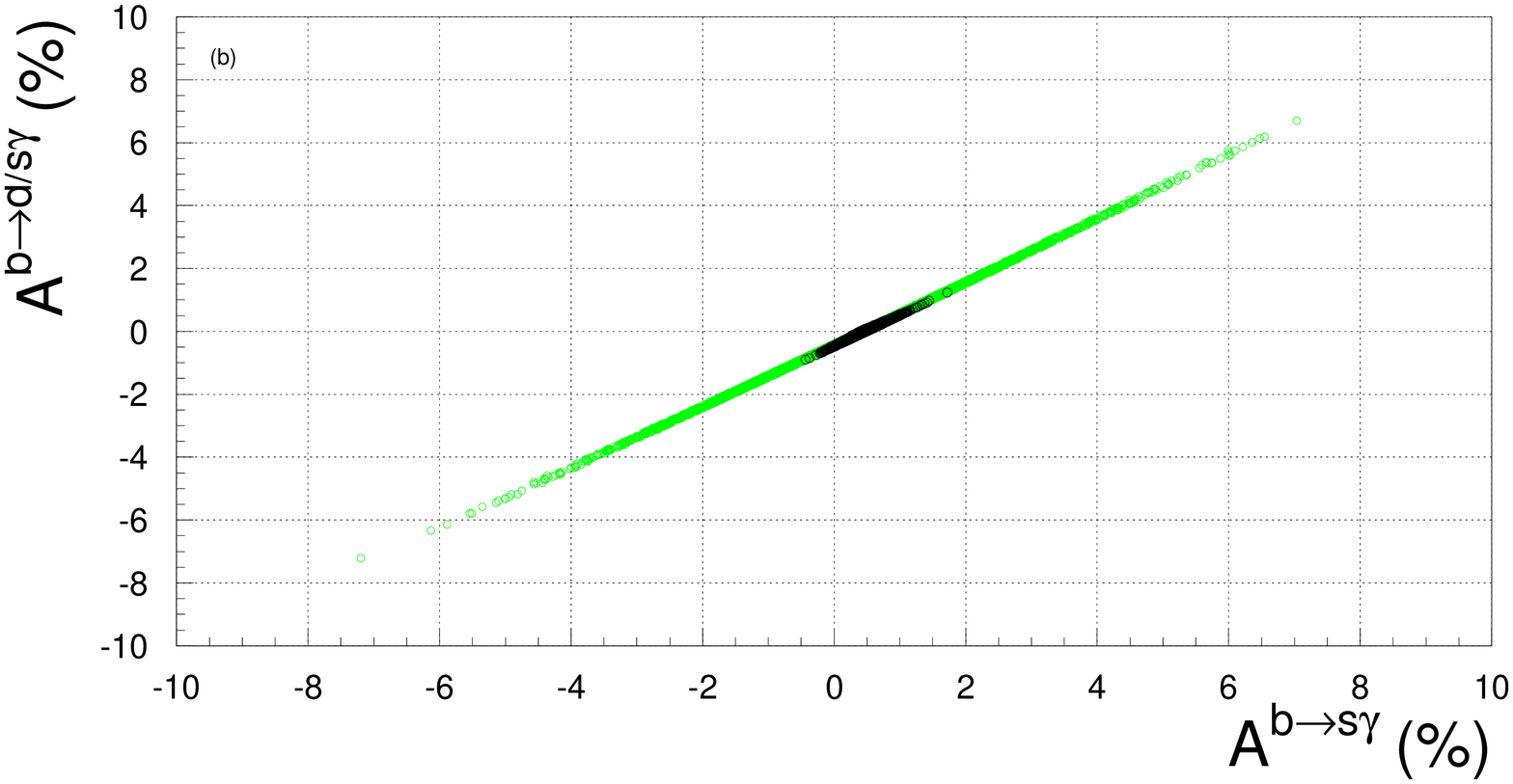,width=.8\linewidth}
\end{center}
\vspace{-1.8cm}
\begin{center}
\epsfig{figure=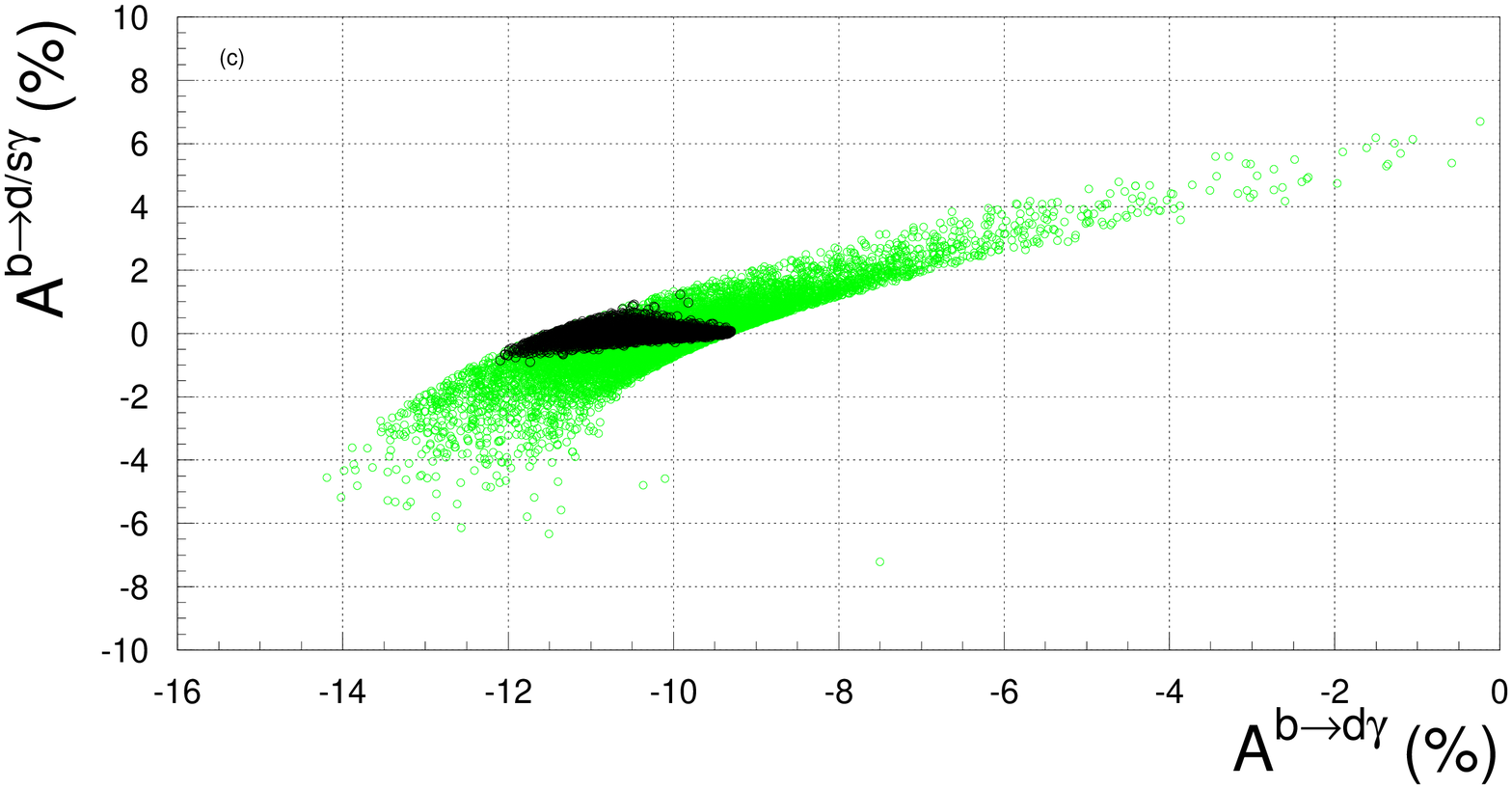,width=.8\linewidth}
\end{center}
\vspace{-1cm}
\caption{Untagged rate asymmetry in the
MFV scenario with non-vanishing flavour-blind phases. The
EDM constraint is relaxed for the green  points and 
imposed on the black  ones.}
\label{MFVwith}
\end{figure}

Scanning the parameter space of the MFV scenario without extra
flavour-blind phases as described above, we find that the untagged CP
asymmetry is completely unaffected. From Fig.~\ref{real} we see that
there are only tiny effects at the $0.02 \%$ level. This shows the
stability of the strict MFV concept in running from the GUT scale down
to the electroweak scale, via fourteen orders of magnitude, due to the
hierarchical structure of the flavour parameters.  We also note that
the tagged $b\to s$ CP asymmetry allows for a bigger ($ \sim \pm
0.2\%$) but still unobservable deviation from the SM value.

Allowing for extra flavour-blind phases for the $A$- and $\mu$
parameters at the GUT scale, significant larger effects in the
untagged asymmetry are possible, as can be seen in Fig.~\ref{MFVwith}
where we present the results of the parameter scanning. In all the
plots, the black (green) points have been obtained with (without)
requiring the EDMs constraint.
In Fig.~\ref{MFVwith}a we plot the untagged asymmetry as a function of
$B(b\to s \gamma)$. One clearly sees that possible effects are still
much below the $5 \%$ threshold, in particular if the EDM constraint
is imposed.
Therefore, a clear discrimination between minimal and general flavour
models is still possible via the untagged CP asymmetry as we will
explore more concretely in the next section.
In Fig.~\ref{MFVwith}b, we show the strong correlation between the
untagged and the tagged $b\to s$ CP asymmetry.  Supersymmetric
contributions to the magnetic and chromo-magnetic Wilson coefficients
are complex, but identical for the $s$ and $d$ sectors: $C^d_{7} =
C^s_{7}$ and $C^d_{8} = C^s_{8}$ to a very high precision. These
relations then imply a strong correlation between the new physics
contributions to the normalized CP asymmetries. They also imply that
the ratio $R_{ds}$ does not deviate appreciably from its SM value:
$R_{ds} \approx R_{ds}^{SM}$. This implies linear proportionality
between tagged and untagged CP asymmetries:
\beq
A_{\rm CP} (B \to X_{d/s} \gamma)_{\rm flavour blind} \sim 
{A_{\rm CP} (B \to X_s \gamma) \over 1 + R_{ds}} \sim
A_{\rm CP} (B \to X_s \gamma) \; .
\eeq
The scatter plot in Fig.~\ref{MFVwith}c shows no direct correlation of
the untagged CP asymmetry to the tagged one in the $b \rightarrow d$
mode.

We conclude that, with respect to a minimal flavour-violating
scenario, there is no theoretical reason to make an extra effort to
measure the tagged CP asymmetry in the $b \rightarrow s$ mode, if the
untagged CP asymmetry is measured: the untagged measurement represents
the cleaner test of the SM. We also stress that a very high precision
would be needed in order to separate MFV scenarios with and without
flavour-blind phases by a measurement of the untagged CP asymmetry.

\section{Model-independent analysis of CP asymmetries}

The model-independent formulae of the CP asymmetries
Eqs.~(\ref{brnum}) and (\ref{acpnum}), based on the operator basis of
Eq.~(\ref{operatorbasis}), allow us to analyze new physics scenarios
considerably more general than the minimal flavour-violating
MSSM. Taking into account the experimental bounds on the $\bar B
\rightarrow X_s \gamma$ branching ratio and CP asymmetry given in
Eqs.~(\ref{world}) and (\ref{acpexp}), we investigate the untagged CP
asymmetry. In particular, we analyze possible correlations between the
tagged and the untagged measurements and their indirect sensitivity to
the CP asymmetry in the $\bar B \rightarrow X_d \gamma$ mode, which
will not be directly measurable in the near future.

Within this model-independent analysis, the $b\to s$ or $b\to d$
sectors are uncorrelated and described by the Wilson coefficients
$C_{7,8}^s$ and $C_{7,8}^d$, respectively. In the following we study
two distinct scenarios in which either $C_{7,8}^s$ or $C_{7,8}^d$ are
allowed to differ from their SM values. These scenarios are very
different from any minimal flavour-violating model (not necessarily
within a supersymmetric framework) in which the $d$ and $s$ sectors
are always correlated: $C_{7,8}^s \buildrel {\rm MFV} \over {=}
C_{7,8}^d$.
\begin{figure}
\begin{center}
\epsfig{figure=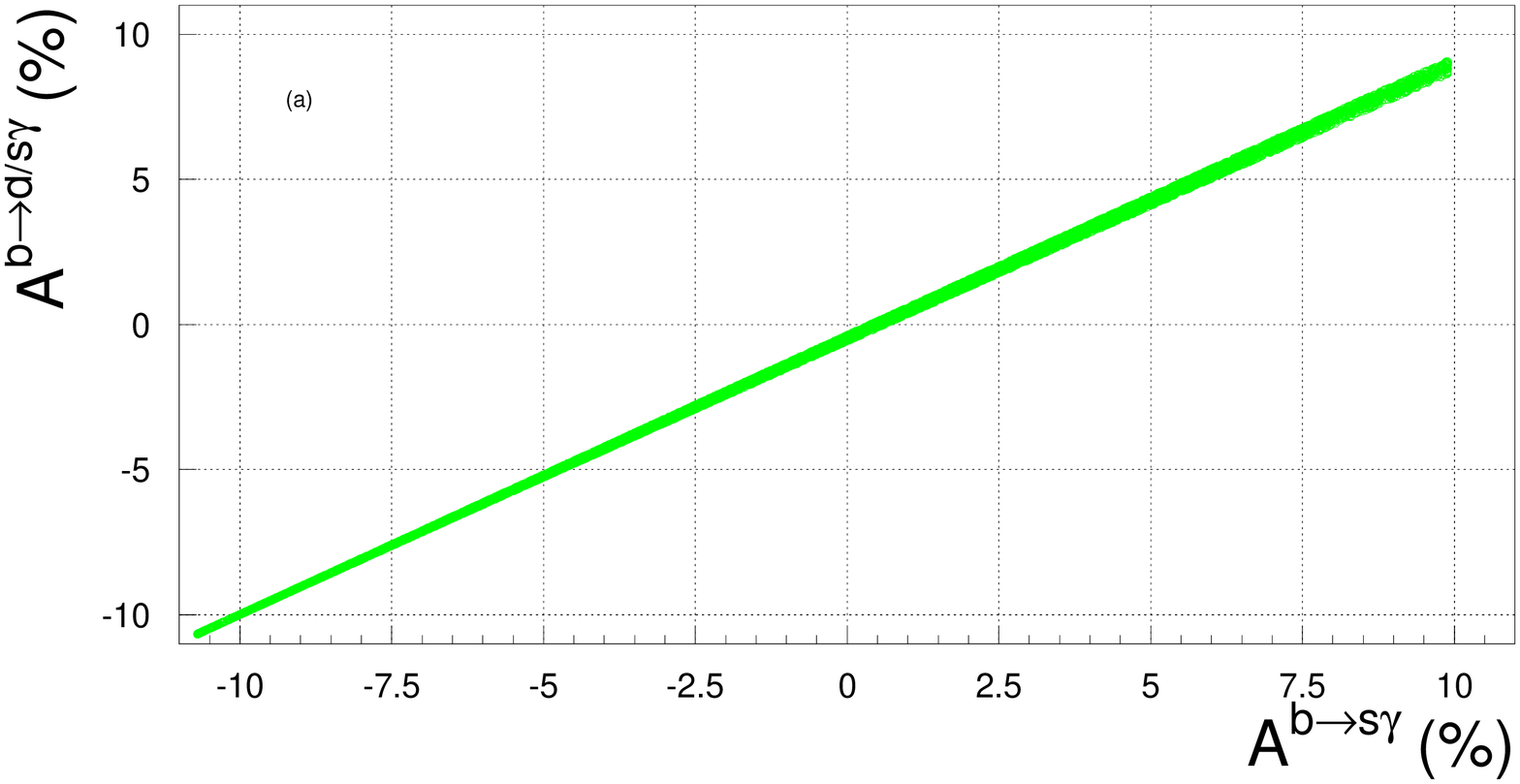,width=.8\linewidth}
\end{center}
\vspace{-1.8cm}
\begin{center}
\epsfig{figure=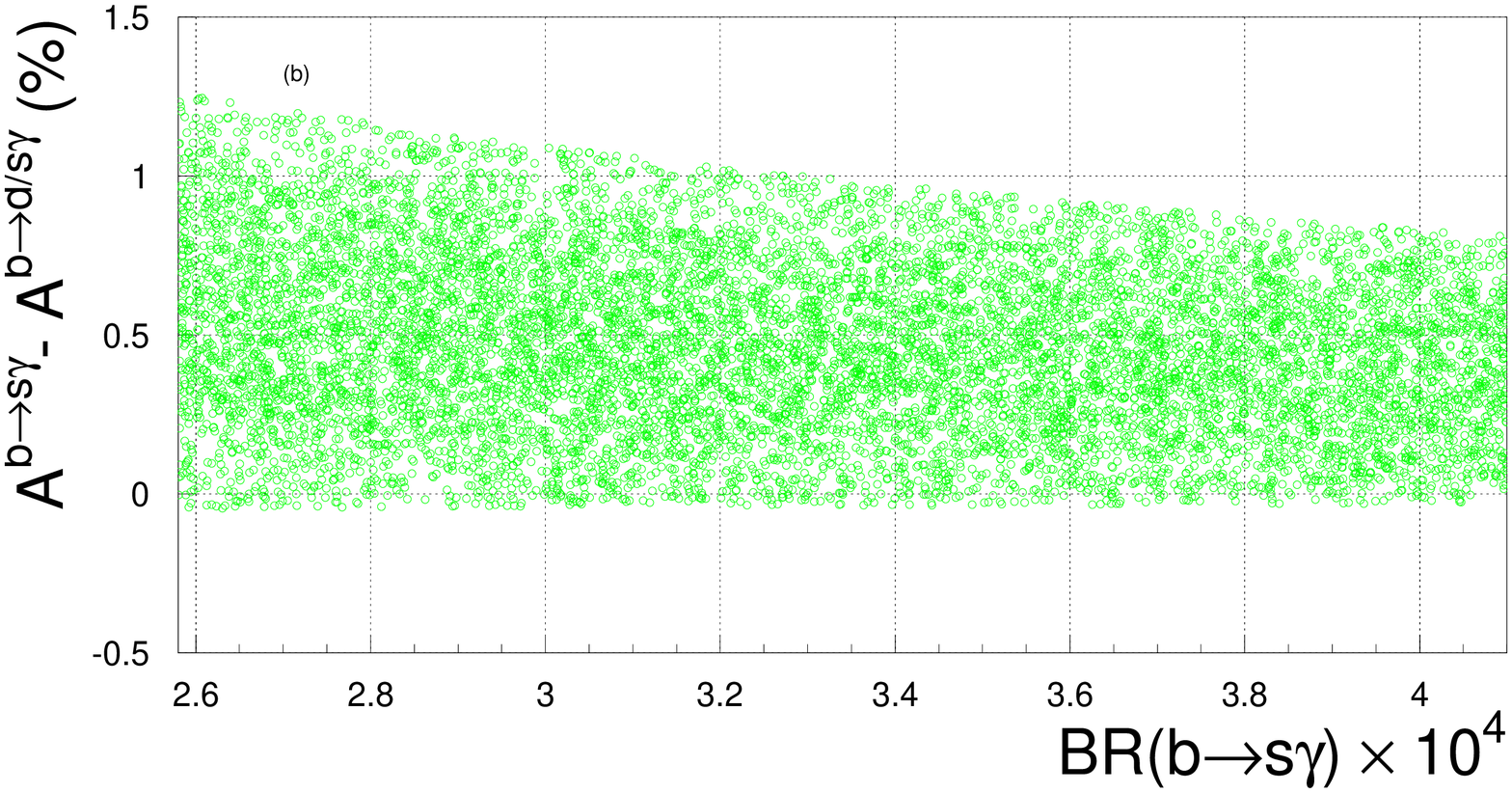,width=.8\linewidth}
\end{center}
\vspace{-.8cm}
\caption{Model-independent analysis with new physics in
$C_{7,8}^s$. Correlation between the tagged and untagged CP
asymmetries.}
\label{C^s}
\end{figure}

We summarize the results of the analysis of the scenario with new
physics in $C_{7,8}^s$ in the scatter plots presented in
Fig.~\ref{C^s}. The points are generated varying $C_{7,8}^s (\mu_0)$
in the complex plane and imposing the experimental constraints $\bar B
\to X_s \gamma $ and $\bar B \to X_s g$. The latter, in particular,
provides the following loose constraint onto $C_8^s$~\cite{AGHL}:
$|C_8^s (\mu_0) /C_8^{s,{\rm SM}} (\mu_0)| < 10$.

In this scenario, the $\bar B \to X_s \gamma$ CP asymmetry receives
large contributions and, indeed, it saturates the experimental bound
given in Eq.~(\ref{acpexp2}). Since the $b\to d$ sector is unaffected,
we expect a strict proportionality between the tagged and untagged CP
asymmetries, as can be seen from Fig.~\ref{C^s}a. In Fig.~\ref{C^s}b
figure, we show in detail the difference between the tagged and
untagged CP asymmetries as a function of the $\bar B \rightarrow X_s
\gamma$ branching ratio. We see that the difference between the two
asymmetries is always below $1.3\%$. Note, finally, that the
difference between the two CP asymmetries is always positive; in fact,
\beq
A_{\rm CP}^{s} - A_{\rm CP}^{d/s} 
= 
{ A_{\rm CP}^{s} (1-R_{ds}) - A_{\rm CP}^{d} R_{ds} \over 1+R_{ds}} 
\simeq 
- {R_{ds} \over 1+R_{ds}} A_{\rm CP}^{d,{\rm SM}} > 0 \,.
\eeq

\begin{figure}
\begin{center}
\epsfig{figure=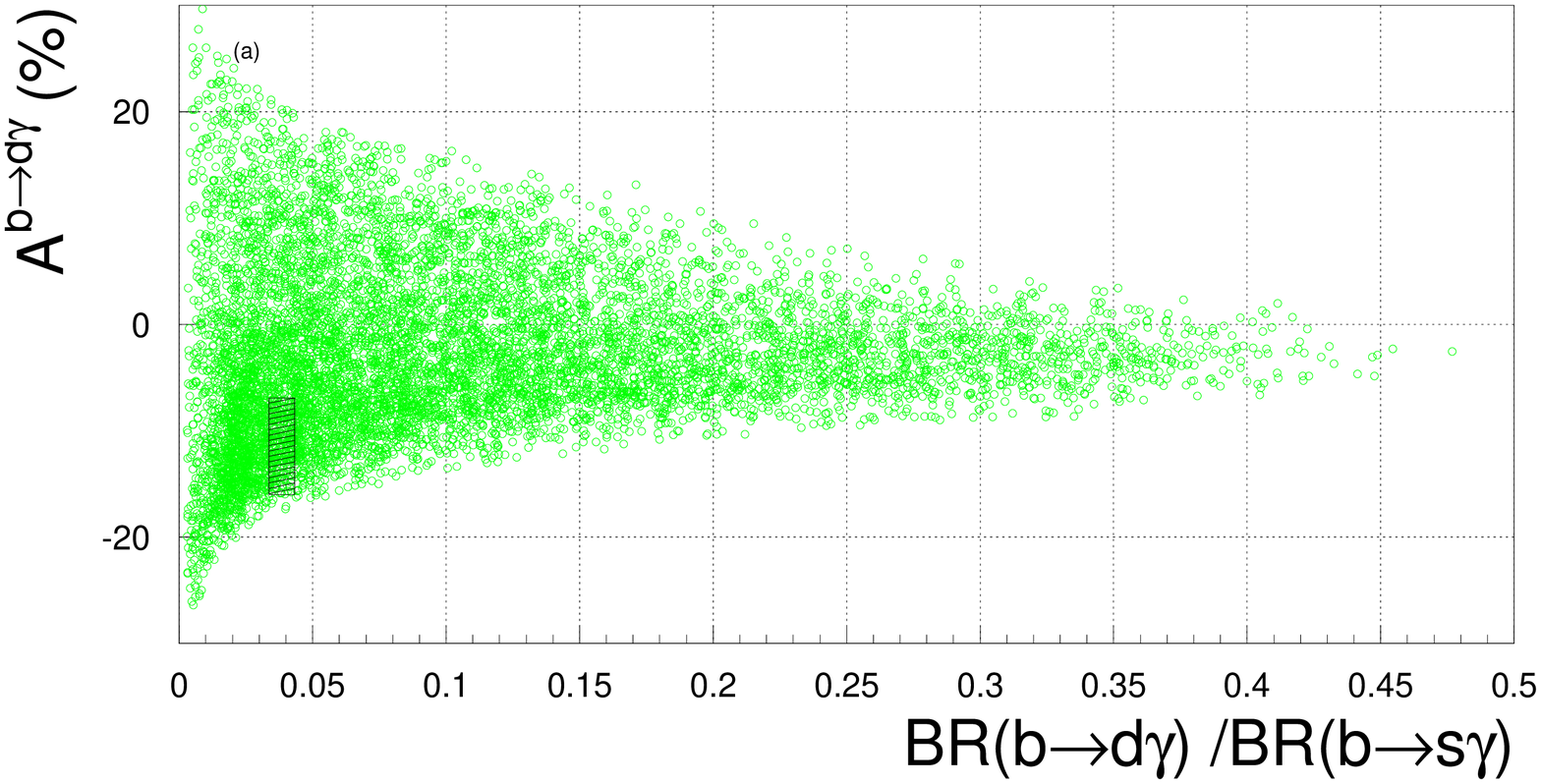,width=.8\linewidth}
\end{center}
\vspace{-1.8cm}
\begin{center}
\epsfig{figure=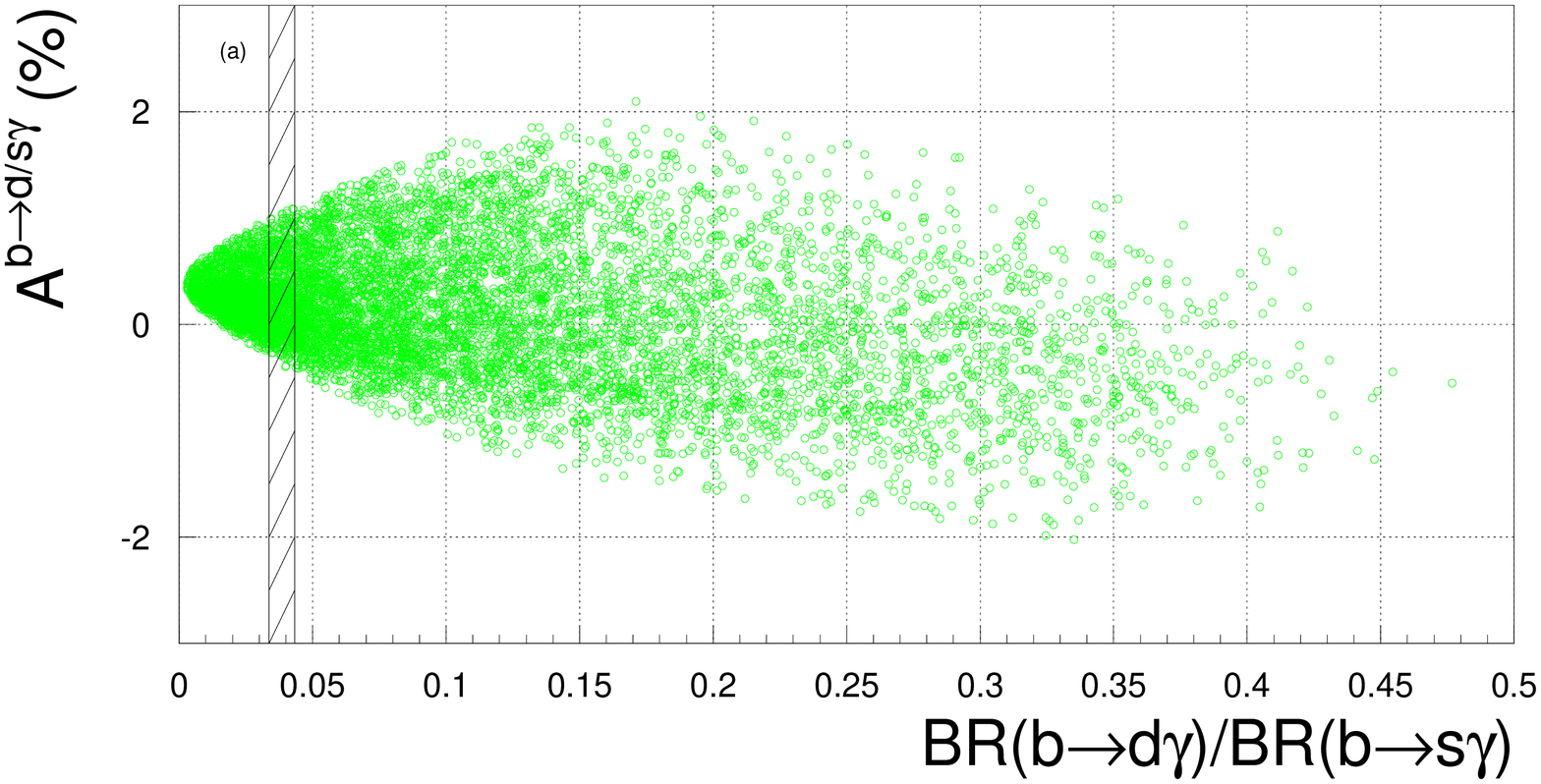,width=.8\linewidth}
\end{center}
\vspace{-1.8cm}
\begin{center}
\epsfig{figure=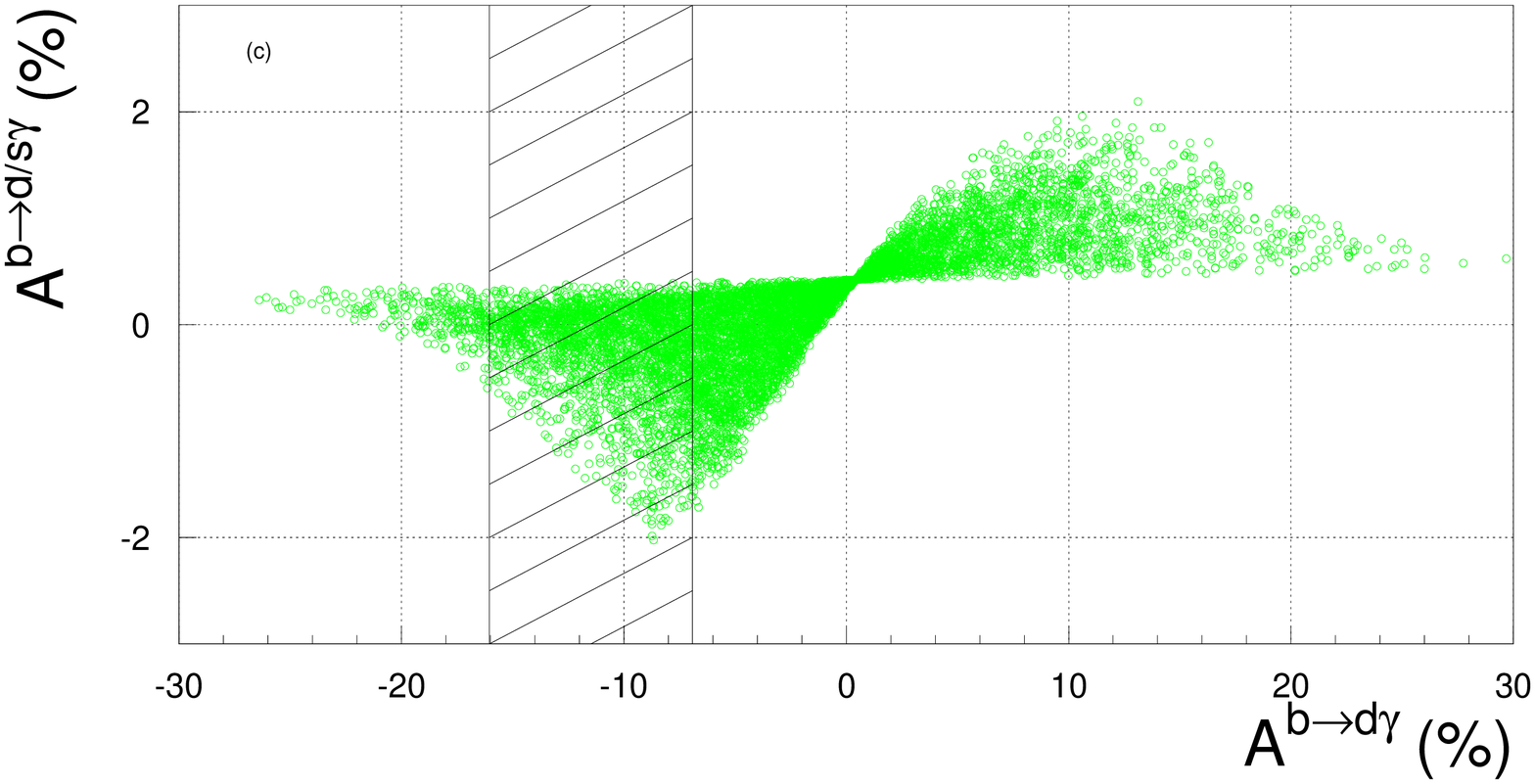,width=.8\linewidth}
\end{center}
\vspace{-1cm}
\caption{Model-independent analysis with new physics in $C_{7,8}^d$.
Correlation between the branching ratio and CP asymmetry in $B\to X_d
\gamma$ and the untagged CP asymmetry. The shaded areas corresponds to
the SM prediction.}
\label{C^d}
\end{figure}

In the second scenario, new physics is present only in the $d$ sector
and we are able to explore the sensitivity of the untagged CP
asymmetry to possible novel effects in the $b \rightarrow d$ mode.
From Fig.~\ref{C^d}, it is clear that such sensitivity is very
restricted; even for very large new physics effects in the $b\to d$ CP
asymmetry or branching ratio, the untagged asymmetry does not exceed
values of $2\%$. In each figure, the shaded band represents the SM
predictions for $A_{\rm CP} (B\to X_d \gamma)$ and for the ratio
$R_{ds}$.
Figure~\ref{C^d}a shows the correlation between branching ratio and CP
asymmetry in $\bar B\to X_d \gamma$. In Figs.~\ref{C^d}b and
\ref{C^d}c, we illustrate the correlation between the $B\to X_d
\gamma$ branching ratio and the untagged CP asymmetry and between the
tagged and the untagged CP asymmetry respectively.

We cross-checked our model-independent analysis in a general flavour
violation scenario within supersymmetry, using the mass insertion
method.  The mass insertion approximation is a well-known useful tool
to study the effect of the large number of flavour-changing parameters
present in the MSSM. The idea is to move into a basis in which the
Yukawas are diagonal without introducing relative rotations between
particles and the corresponding superpartners (i.e. we consider rigid
superfield transformations). In this so--called super-CKM basis, the
squark mass matrices are non-diagonal and represent generic new
sources of flavour-changing neutral currents within the MSSM. We can
then expand the physical amplitudes in powers of these off-diagonal
elements, assuming that they are small with respect to the diagonal
entries. A given process is thus dominated by only few of these mass
insertions.  A comprehensive analysis of all the insertions for the CP
asymmetries in the spirit of the analysis \cite{borzumati,besmer}) is
beyond the scope of the present analysis.  However, we analyzed gluino
contributions with non-vanishing $(m^2_{LR})^d_{23}$ or
$(m^2_{LR})^d_{13}$ and their chiral analogue.  We always consider
mass insertions normalized to the average down squarks mass, i.e.
\beq
(\d^d_{LR})_{23} = {(m^2_{LR})^d_{23} \over {\tilde m}^2} \;\;\;\; {\rm and} \;\;\;\;
(\d^d_{LR})_{13} = {(m^2_{LR})^d_{13} \over {\tilde m}^2} \; ,
\eeq
where we choose $\tilde m \sim 500\, \gev$. Moreover, we have taken
into account bounds on the mass insertions induced by all the various
experimental constraints (see Ref.~\cite{massinsertionsconstraints}
for a detailed description). In the two separate scenarios in which
either $(\d^d_{LR})_{23}\neq 0$ or $(\d^d_{LR})_{13}\neq 0$, we have
not found any significant deviation from the model-independent results
presented above. 


\section{Conclusions}
We have presented updated SM predictions for the branching ratios
$\bar{B} \to X_q \gamma$ ($q=s,d$) and the corresponding CP
asymmetries together with model-independent formulae that can be used
to study the impact of generic new physics interactions on these
observables. We have shown that the untagged CP asymmetry, i.e.~the CP
asymmetry in the $\bar B\to X_{s+d} \gamma$ mode, is extremely
sensitive to new physics contributions. In the SM, in fact, this
observable is negligibly small thanks to U-spin relations and to the
unitarity of the CKM matrix and allows for a clean test, whether
additional CP phases are present or not.

Using the model-independent formulae, we have analysed the untagged CP
asymmetry in several scenarios beyond the SM. We considered the MSSM
with minimal flavour violation and a model with generic contributions
to the Wilson coefficients $C_7$ and $C_8$.

MFV models are characterized by the requirement of expressing all
flavour-changing interactions in terms of powers of the Yukawa
matrices. In a first stage, we assumed the CKM phase to be the only CP
phase present at the grand unification scale. In this restricted
scenario we find that the untagged CP asymmetry receives only very
small contributions: this class of models cannot be distinguished from
the SM with the help of this observable. Subsequently, we allowed the
$\mu$ and $A$ parameters to be complex. After the EDM bounds are taken
into account, only asymmetries below the $2\%$ level survive and we
find a strict proportionality between the untagged ($\bar B\to X_{s+d}
\gamma$) and tagged ($\bar B\to X_{s} \gamma$) CP asymmetries. The
task of distinguishing these two MFV scenarios is beyond the
possibilities of the existing $B$-factories but should be within the
reach of future experiments.

In the model-independent approach, we have allowed for new physics
contributions to the $s$ and $d$ sectors independently. In the first
case, the untagged CP asymmetry can be as large as $\pm 10 \%$, once
the recent experimental data from Belle on the CP asymmetries are
taken into account; more importantly, we found that the tagged and
untagged asymmetries are again strictly proportional to each other. In
the second case with new physics in the $d$ sector, we have not found
untagged CP asymmetries larger than $2 \%$: this implies that the
untagged CP asymmetry is not really sensitive for new physics effects
in the $d$ sector.

With the expected experimental accuracy of $\pm 3 \%$ at the 
$B$-factories, a clear distinction between a minimal and a more general
flavour model is possible through the untagged CP asymmetry.

\section*{Acknowledgments}
We thank Mikolaj Misiak, Daniel Wyler and Gino Isidori for interesting
discussions and suggestions. This work is supported by the Swiss
'Nationalfonds' and partly by the EC-Contract HPRN-CT-2002-00311
(EURIDICE).  Fonds. W.P.~is supported by the Erwin Schr\"odinger
fellowship no.~J2272 of the `Fonds zur F\"orderung der
wissenschaftlichen Forschung' of Austria.

\appendix
\section*{Appendix}
\label{app1}
We collect the explicit expressions for many of the quantities
introduced in sect. 2:

\begin{itemize}
\item 
The ratio of the $\overline{\rm MS}$ running mass of the bottom quark
($m_b^{\overline{\rm MS}} (\mu_0)$) to the 1S mass ($m_b^{1S}$)
is~\cite{GM}~($\mu_0=m_t$):
\beq
r(\mu_0) = 0.578 \left( \alpha_s (M_Z) \over 0.1185 \right)^{-1.0} 
             \left( m_b^{1S} \over 4.69 \right)^{0.23} 
             \left( m_c(m_c) \over 1.25 \right)^{-0.003} 
             \left( \mu_0 \over 165 \right)^{-0.08} 
             \left( \mu_b \over 4.69 \right)^{0.006} \,.
\eeq

\item 
The charm contribution of the perturbative part is given by :
\bea
K_c^{(0)} &=& \sum_{k=1}^8 \eta^{a_k} d_k \,,\\ 
K_c^{(11)} &=& {\alpha_s (\mu_b) \over 4 \pi} \sum_{k=1}^8 \eta^{a_k} 
\Bigg[ 2 \beta_0 a_k d_k \left( \log {m_b \over \mu_b} + \eta \log {\mu_0 \over m_W} \right)
+ \tilde d_k + \tilde d_k^\eta \eta \nn\\
 & & + {\rm Re}\left[ \tilde d_k^a a(z) +  \tilde d_k^b b(z)\right] \,,\\ 
K_c^{(12)} &=& {\alpha_s (\mu_b) \over 4 \pi} \sum_{k=1}^8 \eta^{a_k} \Bigg[ \tilde d_k^{i\pi} \pi
               + {\rm Im}\left[ \tilde d_k^a a(z) +  \tilde d_k^b b(z)\right] \Bigg] \,,\\
K_c^{(13)} &=& {\alpha_s (\mu_b) \over 4 \pi} \sum_{k=1}^8 \eta^{a_k} 
               {\rm Re}\left[ \tilde d_k^a a(z) +  \tilde d_k^b b(z)\right] \,,\\
K_c^{(14)} &=& {\alpha_s (\mu_b) \over 4 \pi} \sum_{k=1}^8 \eta^{a_k} 
               {\rm Im}\left[ \tilde d_k^a a(z) +  \tilde d_k^b b(z)\right] \,,
\eea
where $\eta = \alpha_s (\mu_0) / \alpha_s (\mu_b)$, $\beta_0 = 23/3$,
$z=(m_c/m_b)^2$; the magic numbers $a_k$, $d_k$, $\tilde d_k$, $\tilde
d_k^{a}$, $\tilde d_k^{b}$, $\tilde d_k^{i \pi}$ can be found in
Table~2 of Ref.~\cite{BCMU} and the functions $a(z)$, $b(z)$ are
presented in Appendix~D of Ref.~\cite{GM}.

\item 
The leading order top contribution is:
\bea
K_t^{(01)} &=&  \eta^{4\over 23} {23\over 36} -{8\over 3} (\eta^{4\over 23}-\eta^{2\over 23}) {1\over 3}\,,\\
K_t^{(02)} &=&  \eta^{4\over 23} \,,\\ 
K_t^{(03)} &=&  -{8\over 3} (\eta^{4\over 23}-\eta^{2\over 23}) \,.
\eea

\item 
The next-to-leading order top contribution is:
\bea
\label{kt11}
K_t^{(11)} &=&  
   - {2\over 9} \alpha_s (m_b)^2 
         \Big( \eta^{4\over 23} {23\over 36} -{8\over 3} (\eta^{4\over 23}-\eta^{2\over 23}) {1\over 3} \Big) \nn\\ 
 & &  +{\alpha_s(\mu_0) \over \pi} \log {\mu_0 \over m_t} 4 x {\partial \over \partial x} 
 \left[ -{1\over 2}\eta^{4\over 23} A_0^t (x_t) + {4\over 3} (\eta^{4\over 23}-\eta^{2\over 23}) F_0^t(x_t) \right] \nn\\
&& + {\alpha_s(\mu_b) \over 4 \pi} \Bigg\{ E_0^t (x_t) \sum_{k=1}^8 e_k \eta^{a_k+ {11\over 23}} 
-2 \eta^{4\over 23} \Bigg[ {1\over 4} \eta A_1^t(x_t) +  \Bigg( {12523\over 3174} -{7411\over 4761} \eta -{2\over 9} \pi^2 \nn \\ 
&& -{4\over 3} \left( \log {m_b \over \mu_b} + \eta \log {\mu_0 \over m_t} \right) \Bigg) {23\over 36}
- {2\over 3} \eta F_1^t(x_t) + \Bigg( -{50092 \over 4761} +{1110842 \over 357075} \eta \nn \\
&& +{16\over 27} \pi^2 + {32\over 9} \left( \log {m_b \over \mu_b} + \eta \log {\mu_0 \over m_t} \right) \Bigg)
 {1\over 3} \Bigg] -2 \eta^{2\over 23 } \Bigg[ {2\over 3} \eta F_1^t(x_t) \nn\\
&& +\Bigg( {2745458 \over 357075} - {38890 \over 14283} \eta - {4\over 9} \pi^2 - {16\over 9}  
\left( \log {m_b \over \mu_b} + \eta \log {\mu_0 \over m_t} \right) \Bigg) {1\over 3} \,,\\
K_t^{(12)} &=& - {2\over 9} \alpha_s (m_b)^2  \eta^{4\over 23} \nn\\
& & - {\alpha_s(\mu_b) \over 2 \pi} \eta^{4\over 23} \Bigg( {12523\over 3174} -{7411\over 4761} \eta -{2\over 9} \pi^2 
-{4\over 3} \left( \log {m_b \over \mu_b} + \eta \log {\mu_0 \over m_t} \right) \Bigg) \,, \\
K_t^{(13)} &=&  {16\over 27} \alpha_s (m_b)^2 (\eta^{4\over 23}-\eta^{2\over 23}) 
                 - {\alpha_s(\mu_b) \over 2 \pi} \Bigg\{  \nn\\  
& & \eta^{4\over 23} \Bigg( -{50092 \over 4761} +{1110842 \over 357075} \eta 
+{16\over 27} \pi^2 + {32\over 9} \left( \log {m_b \over \mu_b} + \eta \log {\mu_0 \over m_t} \right) \Bigg) \nn\\
& & +\eta^{2\over 23 } 
  \Bigg( {2745458 \over 357075} - {38890 \over 14283} \eta - {4\over 9} \pi^2 - {16\over 9}  
\left( \log {m_b \over \mu_b} + \eta \log {\mu_0 \over m_t} \right) \Bigg) \Bigg\}  \,,\\
K_t^{(14)} &=& {2\alpha_s(\mu_b) \over 27 }  \eta^{2\over 23 } \,,\\
K_t^{(15)} &=& {2\alpha_s(\mu_b) \over 9 }  \eta^{2\over 23 } \,,
\eea
where the numbers $e_k$ and the functions $A_1^t$ and $F_1^t$ are
given in Ref.~\cite{GM}.

\item 
The electroweak contributions are~\cite{sirlin,baranow,GH1,GH2}:
\bea
\varepsilon^{(11)} & = & {\alpha_{\rm em} (m_Z) \over \alpha_s (\mu_b)} 
\left( {88\over 575 } \eta^{16\over 23} - {40\over 69 } \eta^{-{7\over 23}}+ {32\over 75 } \eta^{-{9\over 23}} 
\right) -  {\alpha_{\rm em} (m_Z) \over \pi} \log {m_Z\over \mu_b} r (\mu_0) \eta^{4\over 23}  \,,\\
\varepsilon^{(12)} & = & {\alpha_{\rm em} (m_Z) \over \alpha_s (\mu_b)} 
\left( -{704\over 1725 } \eta^{16\over 23} + {640\over 1449 } \eta^{{14\over 23}}+ {32\over 1449 } \eta^{-{7\over 23}} 
-{32\over 575} \eta^{-{9\over 23}} \right) \nn\\
& & - {\alpha_{\rm em} (m_Z) \over \pi} \log {m_Z\over \mu_b} r (\mu_0) {8\over 3} 
    (\eta^{2\over 23}-\eta^{4\over 23})  \,.
\eea

\item 
Finally, the bremsstrahlung functions $\phi (\delta,z)$ appearing in
the expression for $B(E_0)$ in Eq.~(\ref{be0}) coincide with the
functions given in Appendix E of Ref.~\cite{GM} with the only
exception of $\phi_{27}$ (and consequently also of $\phi_{17}$,
$\phi_{28}$ and $\phi_{18}$) which has to be replaced by
\bea
\phi_{27} (\delta) = -{8 z^2\over 9} \left[ 
\delta \int_0^{(1-\delta)/z} {\rm d}t \; \left( G(t) +{t\over 2} \right) 
+ \int_{(1-\delta)/z}^{1/z} {\rm d}t \; (1- zt)  \left( G(t) +{t\over 2} \right)
\right] \,,
\eea
where $G(t)$ is given in Eq. (E.8) of Ref.~\cite{GM}.

\item 
The functions in Eq. (\ref{acp1}) are given by 
\bea
\Delta \Gamma_{q g \gamma} 
& = & 
{2 \alpha_s (\mu_b) \over \pi } \left( C_2 (\mu_b) - {1\over 6}  C_1 (\mu_b) \right) {\rm Im} [\phi_{27}  (\delta)] \,
{\rm Im} \left[  \left( C_7 (\mu_b) - {1\over 3}  C_8 (\mu_b) \right) (1 + \epsilon_q^*) \right] \,  \nn \\ && \\
\Delta \Gamma_{q \gamma}  
& = & 2\, {\rm Im} \Big[ -  K_c^{(0)}  K_c^{(14)} \epsilon_q  
+ r (\mu_0)\; \Big( K_t^{(0)}  K_t^{(1)i*} + K_c^{(14)}  K_t^{(0)} \epsilon_q^{*} +  K_c^{(12)}  K_t^{(0)}  \nn\\
& &  \hskip 1.1cm -  K_c^{(0)}  K_t^{(1)i} \Big) \Big] \,.  
\eea

In comparing with the results given in the first reference
in~\cite{summarys}, we note that $z b(z, \delta) = 9/(8 \pi){\rm Im}
[\phi_{27} (\delta)]$.

\end{itemize}

\end{document}